\documentclass[graybox,envcountchap,openany]{svmult}

\usepackage{type1cm}         
\usepackage{makeidx}         
\usepackage{wrapfig}
\usepackage{graphicx}  
\usepackage{amsmath}
\numberwithin{equation}{chapter}

\usepackage{multicol}        
\usepackage[bottom]{footmisc}
\usepackage{multirow} 
\usepackage[bottom]{footmisc}
\usepackage{makecell} 
\usepackage{newtxtext}       %
\usepackage[varvw]{newtxmath}       
\usepackage{natbib}
\usepackage{array}
\usepackage{siunitx}
\usepackage{lineno}
\usepackage{setspace}
\usepackage{dirtytalk}
\usepackage[T1]{fontenc}
\usepackage[utf8]{inputenc}
\usepackage[varvw]{newtxmath}       
\usepackage{amsmath}
\usepackage{empheq}
\usepackage{graphicx}
\usepackage{dcolumn}
\usepackage{bm}
\usepackage{graphicx}        
\usepackage{booktabs}        
\makeindex            

\newcommand{\IND}[1]{\index{#1}#1}

\fontsize{12}{14}\selectfont
\sloppypar
\begin{document}
\pagenumbering{roman}
\setcounter{page}{1}


\mainmatter

\newcommand{\op}         {\IM{!open problem}}

\newcommand{\dg}        {$^o$}
\newcommand{\ra}            {$\rightarrow$}
\newcommand{\ueber}   [1]{\vspace{0.4 cm}\NO{\bf #1}\vspace{0.2 cm}}
\newcommand{\INDX}    [1]{#1\protect\index{#1}}
\newcommand{\IM}      [1]{{\em #1}\index{#1}}
\newcommand{\SOI}        {Southern Oscillation Index \index{Southern Oscillation!index}}
\newcommand{\ENSO}          {\INDX{El Ni\~no/Southern Oscillation}}
\newcommand{\SO}        {\INDX{Southern Oscillation}}
\newcommand{\EN}          {\INDX{El Ni\~no}}
\newcommand{\pad}     [2] {\partial_{#2} {#1}}    
\newcommand{\oo}      [1] {\frac{1}{#1}}
\newcommand{\mb}      [1]{\markboth{CHAPTER \thechapter. ~#1}{}}
\newcommand{\BTAB}        {\begin{table}}
\newcommand{\ETAB}        {\end{table}}
\newcommand{\DEF}     [1] {\\[3pt]{\it{#1}}\\[3pt]}
\newcommand{\DEFL}    [1] {\\[3pt]{\it{#1}}}
\newcommand{\SC}      [2]{\section{#1} \label{sec:#2} }
\newcommand{\SSC}     [1]{\subsection{#1}}
\newcommand{\SSSC}    [1]{\subsubsection{#1}}
\newcommand{\PA}      [1]{\paragraph{#1.}}
\newcommand{\LIT}     [1]{\newline{\bf{#1}:}}
\newcommand{\EQ}      [1]{(\ref{#1})}
\newcommand{\SEC}     [1]{\ref{sec:#1}}
\newcommand{\CHA}     [1]{\ref{chap:#1}}
\newcommand{\FIG}     [1]{Figure \ref{#1}}
\newcommand{\TAB}     [1]{Table \ref{#1}}
\newcommand{\BE}         {\begin{equation}}
\newcommand{\BEQ}     [1]{\begin{equation}\label{#1}}
\newcommand{\EE}         {\end{equation}}
\newcommand{\BI}         {\begin{itemize}}
\newcommand{\EI}         {\end{itemize}}
\newcommand{\BA}         {\begin{eqnarray}}
\newcommand{\EA}         {\end{eqnarray}}
\newcommand{\ba}      [1]{\begin{array}{#1}}
\newcommand{\ea}         {\end{array}}
\newcommand{\VEC}     [1]{\!\left(\!\!\!\ba{c} #1 \ea \!\!\!\right)\!\!}
\newcommand{\BANN}{\begin{eqnarray*}}
\newcommand{\EANN}{\end{eqnarray*}}
\newcommand{\BENN} {\begin{displaymath}}
\newcommand{\EENN} {\end{displaymath}}
\newcommand{\CAPTION}   [2] {\caption{\sl{#1}\protect\label{#2}}}
\newcommand{\ABSTRACT}  [1] {\begin{abstract}{\normalsize #1}\end{abstract}}
\newcommand{\LINE} {\begin{center} {------------------------------------------------}\end{center}}
\newcommand{\NN} \nonumber
\newcommand{\EF}          {\end{figure}}
\newcommand{\EFdouble}    {\LINE\end{figure*}}
\newcommand{\EFd}         {\end{figure*}}
\newcommand{\BF}       [2]{\begin{figure}[htb]\CAPTION{#1}{#2}\EF}
\newcommand{\BFdouble} [2]{\begin{figure*}[htb]\LINE\CAPTION{#1}{#2}\EFdouble}
\newcommand{\BFs}      [3]{\begin{figure}[htb]\vspace{#3}\CAPTION{#1}{#2}\EF}
\newcommand{\BFsd}     [3]{\begin{figure*}[htb]\vspace{#3}\CAPTION{#1}{#2}\EFd}
\newcommand{\BFC}      [2]{\begin{figure}\CAPTION{#1}{#2}}
\newcommand{\psplot}   [2]{\begin{minipage}[e] {\textwidth} \centering \includegraphics [width=#1 cm] {#2} \end{minipage}}
\newcommand{\FIGURE}   [4]{\begin{figure}\CAPTION{#1}{#2} \psplot{#3}{#4}}
\newcommand{\FIGUREd}  [4]{\begin{figure*}\CAPTION{#1}{#2}\psplot{#3}{#4}}

\newcommand{\TITEM} {\item [$\bigodot$]}
\newcommand{\NO}   {\noindent}
\newcommand{\NL}   {\newline\NO}
\newcommand{\SIGMA} {\mbox{${\bf\Sigma}$}}
\newcommand{\TBC} {{\em --- to be continued ---}\newline}
\newcommand{\cotwo}   {\mbox{CO$_2$}}
\newcommand{\ESIGMA} {\mbox{${\bf\hat\Sigma}$}}
\newcommand{\VAR}     [1]{\mbox{{\sc Var}$\left({#1}\right)$}}
\newcommand{\COV}     [2]{\mbox{{\sc Cov}$\left({#1},{#2}\right)$}}
\newcommand{\EXP}     [1]{\mbox{{\bf{\sc E}}$\left({#1}\right)$}}
\newcommand{\NORM} [1] {\mbox{$\parallel {#1} \parallel$}}
\newcommand{\DOTP} [2] {\mbox{$\langle {#1},{#2} \rangle$}}
\newcommand{\PTILDE} {\mbox{$\tilde{\sl p}$}}
\newcommand{\PROB} [1] {\mbox{${\sl prob}\left({#1}\right)$}}
\newcommand{\CRIN}  {\mbox{${\kappa}_{\PTILDE}$}}
\newcommand{\EVH} [1] {\mbox{$\hat{{\vec e}^{#1}}$}}
\newcommand{\EV} [1] {\mbox{${\vec e}^{#1}$}}
\newcommand{\PV} [1] {\mbox{${\vec p}\:^{#1}$}}
\newcommand{\QV}      [1]{\mbox{${\vec q}\:^{#1}$}}
\newcommand{\Pv}      [2]{\mbox{${\vec p}_{#2}^{\,#1}$}}
\newcommand{\Ev}      [2]{\mbox{${\vec e}_{#2}^{\,#1}$}}
\newcommand{\PVA}     [1]{\mbox{$\Pv{#1}{A}$}}
\newcommand{\PVR}     [1]{\mbox{$\Pv{#1}{R}$}}
\newcommand{\PVH} [1] {\mbox{${\hat{\vec p}\:^{#1}}$}}
\newcommand{\EAL}  {\mbox{$\hat{\alpha}$}}
\newcommand{\al}      [2]{\mbox{${\alpha}_{#1}^{\,#2}$}}
\newcommand{\ELAM}  {\mbox{$\hat{\lambda}$}}
\newcommand{\RVa}  {{\bf a}\/}
\newcommand{\RVA}  {{\bf A}\/}
\newcommand{\RVva} {\mbox{${\vec\RVa}$}}
\newcommand{\RVVA} {\mbox{${\vec\RVA}$}}
\newcommand{\RVb}  {{\bf b}}
\newcommand{\RVB}  {{\bf B}}
\newcommand{\RVvb} {\mbox{${\vec\RVb}$}}
\newcommand{\RVVB} {\mbox{${\vec\RVB}$}}
\newcommand{\RVc}  {{\bf c}}
\newcommand{\RVC}  {{\bf C}}
\newcommand{\RVvc} {\mbox{${\vec\RVc}$}}
\newcommand{\RVVC} {\mbox{${\vec\RVC}$}}
\newcommand{\RVd}  {{\bf d}}
\newcommand{\RVD}  {{\bf D}}
\newcommand{\RVvd} {\mbox{${\vec\RVd}$}}
\newcommand{\RVVD} {\mbox{${\vec\RVD}$}}
\newcommand{\RVe}  {{\bf e}}
\newcommand{\RVE}  {{\bf E}}
\newcommand{\RVve} {\mbox{${\vec\RVe}$}}
\newcommand{\RVVE} {\mbox{${\vec\RVE}$}}
\newcommand{\RVf}  {{\bf f}}
\newcommand{\RVF}  {{\bf F}}
\newcommand{\RVvf} {\mbox{${\vec\RVf}$}}
\newcommand{\RVVF} {\mbox{${\vec\RVF}$}}
\newcommand{\RVg}  {{\bf g}}
\newcommand{\RVG}  {{\bf G}}
\newcommand{\RVvg} {\mbox{${\vec\RVg}$}}
\newcommand{\RVVG} {\mbox{${\vec\RVG}$}}
\newcommand{\RVh}  {{\bf h}}
\newcommand{\RVH}  {{\bf H}}
\newcommand{\RVvh} {\mbox{${\vec\RVh}$}}
\newcommand{\RVVH} {\mbox{${\vec\RVH}$}}
\newcommand{\RVi}  {{\bf i}}
\newcommand{\RVI}  {{\bf I}}
\newcommand{\RVvi} {\mbox{${\vec\RVi}$}}
\newcommand{\RVVI} {\mbox{${\vec\RVI}$}}
\newcommand{\RVj}  {{\bf j}}
\newcommand{\RVJ}  {{\bf J}}
\newcommand{\RVvj} {\mbox{${\vec\RVj}$}}
\newcommand{\RVVJ} {\mbox{${\vec\RVJ}$}}
\newcommand{\RVk}  {{\bf k}}
\newcommand{\RVK}  {{\bf K}}
\newcommand{\RVvk} {\mbox{${\vec\RVk}$}}
\newcommand{\RVVK} {\mbox{${\vec\RVK}$}}
\newcommand{\RVl}  {{\bf l}}
\newcommand{\RVL}  {{\bf L}}
\newcommand{\RVvl} {\mbox{${\vec\RVl}$}}
\newcommand{\RVVL} {\mbox{${\vec\RVL}$}}
\newcommand{\RVm}  {{\bf m}}
\newcommand{\RVM}  {{\bf M}}
\newcommand{\RVvm} {\mbox{${\vec\RVm}$}}
\newcommand{\RVVM} {\mbox{${\vec\RVM}$}}
\newcommand{\RVn}  {{\bf n}}
\newcommand{\RVN}  {{\bf N}}
\newcommand{\RVvn} {\mbox{${\vec\RVn}$}}
\newcommand{\RVVN} {\mbox{${\vec\RVN}$}}
\newcommand{\RVo}  {{\bf o}}
\newcommand{\RVO}  {{\bf O}}
\newcommand{\RVvo} {\mbox{${\vec\RVo}$}}
\newcommand{\RVVO} {\mbox{${\vec\RVO}$}}
\newcommand{\RVp}  {{\bf p}}
\newcommand{\RVP}  {{\bf P}}
\newcommand{\RVvp} {\mbox{${\vec\RVp}$}}
\newcommand{\RVVP} {\mbox{${\vec\RVP}$}}
\newcommand{\RVq}  {{\bf q}}
\newcommand{\RVQ}  {{\bf Q}}
\newcommand{\RVvq} {\mbox{${\vec\RVq}$}}
\newcommand{\RVVQ} {\mbox{${\vec\RVQ}$}}
\newcommand{\RVr}  {{\bf r}}
\newcommand{\RVR}  {{\bf R}}
\newcommand{\RVvr} {\mbox{${\vec\RVr}$}}
\newcommand{\RVVR} {\mbox{${\vec\RVR}$}}
\newcommand{\RVs}  {{\bf s}}
\newcommand{\RVS}  {{\bf S}}
\newcommand{\RVvs} {\mbox{${\vec\RVs}$}}
\newcommand{\RVVS} {\mbox{${\vec\RVS}$}}
\newcommand{\RVt}  {{\bf t}}
\newcommand{\RVT}  {{\bf T}}
\newcommand{\RVvt} {\mbox{${\vec\RVt}$}}
\newcommand{\RVVT} {\mbox{${\vec\RVT}$}}
\newcommand{\RVu}  {{\bf u}}
\newcommand{\RVU}  {{\bf U}}
\newcommand{\RVvu} {\mbox{${\vec\RVu}$}}
\newcommand{\RVVU} {\mbox{${\vec\RVU}$}}
\newcommand{\RVv}  {{\bf v}}
\newcommand{\RVV}  {{\bf V}}
\newcommand{\RVvv} {\mbox{${\vec\RVv}$}}
\newcommand{\RVVV} {\mbox{${\vec\RVV}$}}
\newcommand{\RVw}  {{\bf w}}
\newcommand{\RVW}  {{\bf W}}
\newcommand{\RVvw} {\mbox{${\vec\RVw}$}}
\newcommand{\RVVW} {\mbox{${\vec\RVW}$}}
\newcommand{\RVx}  {{\bf x}}
\newcommand{\RVX}  {{\bf X}}
\newcommand{\RVvx} {\mbox{${\vec\RVx}$}}
\newcommand{\RVVX} {\mbox{${\vec\RVX}$}}
\newcommand{\RVy}  {{\bf y}}
\newcommand{\RVY}  {{\bf Y}}
\newcommand{\RVvy} {\mbox{${\vec\RVy}$}}
\newcommand{\RVVY} {\mbox{${\vec\RVY}$}}
\newcommand{\RVz}  {{\bf z}}
\newcommand{\RVZ}  {{\bf Z}}
\newcommand{\RVvz} {\mbox{${\vec\RVz}$}}
\newcommand{\RVVZ} {\mbox{${\vec\RVZ}$}}

\newcommand{\matrixx}[1] {\mbox{${\cal{#1}}$}}
\newcommand{\MA}    {\matrixx{A}}
\newcommand{\MB}    {\matrixx{B}}
\newcommand{\MC}    {\matrixx{C}}
\newcommand{\MD}    {\matrixx{D}}
\newcommand{\ME}    {\matrixx{E}}
\newcommand{\MF}    {\matrixx{F}}
\newcommand{\MG}    {\matrixx{G}}
\newcommand{\MH}    {\matrixx{H}}
\newcommand{\MI}    {\matrixx{I}}
\newcommand{\MJ}    {\matrixx{J}}
\newcommand{\MK}    {\matrixx{K}}
\newcommand{\ML}    {\matrixx{L}}
\newcommand{\MM}    {\matrixx{M}}
\newcommand{\MO}    {\matrixx{O}}
\newcommand{\MP}    {\matrixx{P}}
\newcommand{\MQ}    {\matrixx{Q}}
\newcommand{\MR}    {\matrixx{R}}
\newcommand{\MS}    {\matrixx{S}}
\newcommand{\MT}    {\matrixx{T}}
\newcommand{\MU}    {\matrixx{U}}
\newcommand{\MV}    {\matrixx{V}}
\newcommand{\MX}    {\matrixx{X}}
\newcommand{\MY}    {\matrixx{Y}}
\newcommand{\MZ}    {\matrixx{Z}}

\newcommand{\am}          {angular momentum}
\newcommand{\sh}          {Southern Hemisphere}
\newcommand{\oam}         {$\Omega$ angular momentum}
\newcommand{\ama}         {angular momenta}
\newcommand{\qg}          {quasi-geostrophic}
\newcommand{\f}        {{\it f}}
\newcommand{\ept}        {{\varepsilon_{t}}}

\newcommand{\ddt}     [1]{{d #1 \over dt}}

\newcommand{\lat}        {{\varphi}}
\newcommand{\lon}        {{\theta}}
\newcommand{\pp}         {{{\bf p}}}
\newcommand{\rr}         {{{\bf r}}}
\newcommand{\vvp}        {{{\bf v}^{'}}}
\newcommand{\vvpp}       {{{\bf v}^{''}}}

\newcommand{\xvec}{\mbox{\boldmath $x$}}
\newcommand{\uvec}{\mbox{\boldmath $u$}}
\newcommand{\mvec}{\mbox{\boldmath $m$}}
\newcommand{\nvec}{\mbox{\boldmath $n$}}
\newcommand{\qvec}{\mbox{\boldmath $q$}}
\newcommand{\vvec}{\mbox{\boldmath $v$}}
\newcommand{\kvec}{\mbox{\boldmath $k$}}
\newcommand{\zvec}{\mbox{\boldmath $z$}}
\newcommand{\Dvec}{\mbox{\boldmath $D$}}
\newcommand{\Complex}{\mbox{\boldmath {\cal $C$}}}
\newcommand{\Fvec}{\mbox{\boldmath $F$}}
\newcommand{\Ivec}{\mbox{\boldmath $I$}}
\newcommand{\Jvec}{\mbox{\boldmath $J$}}
\newcommand{\Kvec}{\mbox{\boldmath $K$}}
\newcommand{\Nvec}{\mbox{\boldmath $N$}}
\newcommand{\Rvec}{\mbox{\boldmath $R$}}
\newcommand{\Svec}{\mbox{\boldmath $S$}}
\newcommand{\Uvec}{\mbox{\boldmath $U$}}
\newcommand{\Pivec}{\mbox{\boldmath $\Pi$}}
\newcommand{\alphavec}{\mbox{\boldmath $\alpha$}}
\newcommand{\betavec}{\mbox{\boldmath $\beta$}}
\newcommand{\gammavec}{\mbox{\boldmath $\gamma$}}
\newcommand{\deltavec}{\mbox{\boldmath $\delta$}}
\newcommand{\epsilonvec}{\mbox{\boldmath $\epsilon$}}
\newcommand{\lambdavec}{\mbox{\boldmath $\lambda$}}
\newcommand{\psivec}{\mbox{\boldmath $\psi$}}
\newcommand{\Psivec}{\mbox{\boldmath $\Psi$}}
\newcommand{\nablavec}{\mbox{\boldmath $\nabla$}}
\newcommand{\sigmavec}{\mbox{\boldmath $\sigma$}}
\newcommand{\tauvec}{\mbox{\boldmath $\tau$}}
\newcommand{\omegavec}{\mbox{\boldmath $\omega$}}
\newcommand{\Omegavec}{\mbox{\boldmath $\Omega$}}
\newcommand{\tce} {tides and climate}  
\newcommand{\Lte}{\INDX{Laplace tidal equations}}
\newcommand{\swe}{shallow water equations}
\newcommand{\CFL}{Courant-Friedrich-Levy criterion}
\newcommand{\BSH}{Bundesamt f\"ur Seeschiffahrt und Hydrographie}
\newcommand{\SST}{sea surface temperature}
\newcommand{\ECMWF}{European Center for Medium Range Forecasts}
\newcommand{\acc}{\INDX{anomaly correlation coefficient}}
\newcommand{\NMC}{National Meteorological Center}
\newcommand{\NCEP}{\INDX{National Center for Environmental Prediction}}
\newcommand{\IPCC}{\INDX{Intergovernmental Panel on Climate Change}}
\newcommand{\acof}{auto-covariance function\index{auto-covariance function}}
\newcommand{\acrf}{auto-correlation function\index{auto-correlation function}}
\newcommand{\COR}     [2]{\mbox{Cor$\left({#1},{#2}\right)$}}

\def\sew{\hbox{\bf sew}}
\def\unif{\hbox{\rm unif}}
\def\xv{{\rm xv}}

\def\real{$\rm I\!R$}
\def\new{{\rm new}} 
\def\year{{\rm year}}

\def\beq{\begin{eqnarray}}
\def\eeq{\end{eqnarray}}

\def\beqn{\begin{eqnarray*}}  
\def\eeqn{\end{eqnarray*}}

\def\hoppp{\bigskip\noindent}
\def\hopp{\medskip\noindent}
\def\hop{\smallskip\noindent}
\def\E{{\rm E}}
\def\Var{{\rm Var}}
\def\Cov{{\rm Cov}}
\def\cov{{\rm cov}}
\def\dd{{\rm d}}
\def\N{{\rm N}}
\def\lik{{\rm lik}}
\def\qlik{{\rm qlik}}
\def\Pr{P}
\def\pr{{\rm pr}}
\def\calM{{\cal M}}

\def\eigen{{\rm eigen}}
\def\maxeig{{\rm max\,eig}}
\def\mineig{{\rm min\,eig}}
\def\quadandquad{\quad \hbox{and} \quad}
\def\arr{\rightarrow}
\def\hatt{\widehat}
\def\tilda{\widetilde}
\def\emp{{\rm emp}}
\def\sumin{\sum_{i=1}^n}
\def\sumjk{\sum_{j=1}^k}
\def\prodin{\prod_{i=1}^n}
\def\maxin{\max_{i\le n}}
\def\eps{\varepsilon}
\def\half{\hbox{$\oo{2}$}}
\def\third{\hbox{$\oo{3}$}}
\def\quart{\hbox{$\oo{4}$}}
\def\onebyn{n^{-1}}
\def\rootn{\sqrt{n}}
\def\data{{\rm data}}
\def\obs{{\rm obs}}
\def\midd{\,|\,}
\def\tr{{\rm t}}
\def\dell{\partial}
\def\EL{{\rm EL}}
\def\calU{{\cal U}}
\newcommand\calR{{\cal R}}
\def\BB{B}
\def\calT{{\cal T}}
\def\prof{{\rm prof}}
\def\KL{{\rm KL}}

\def\calD{{\cal D}}
\def\one{{\bf 1}}
\def\argmin{{\rm argmin}}
\def\argmax{{\rm argmax}}
\def\Bernstein{{Bernshte\u\i n}}
\def\true{{\rm true}}
\def\const{{\rm const.}}
\def\sd{{\rm sd}}
\def\med{{\rm med}}
\def\RIC{{\rm RIC}}
\def\aic{{\rm aic}}
\def\bic{{\rm bic}}
\def\jic{{\rm jic}}
\def\AIC{{\rm AIC}}
\def\BIC{{\rm BIC}}
\def\WIC{{\rm WIC}}
\def\AWIC{{\rm AWIC}}
\def\BJIC{{\rm BJIC}}
\def\Tr{{\rm Tr}}
\def\pen{{\rm pen}}
\def\mse{{\rm mse}}
\def\simm{{\rm sim}}
\def\RR{{\cal R}}
\def\ML{{\rm ML}}
\def\pois{{\rm Pois}}

\def\cc{{\rm cc}}
\def\Expo{{\rm Expo}}
\def\Dir{{\rm Dir}}
\def\newsquare{{\ \vrule height0.5em width0.5em depth-0.0em}}
\def\ml{{\rm ml}}
\def\fic{{\rm fic}}
\def\nonpara{{\rm np}}
\def\ib{{\rm ib}}
\def\ip{{\rm ip}}
\def\bsq{{\rm bsq}}

\def\corr{{\rm corr}}
\def\diag{{\rm diag}}
\def\epp{{\rm epp}}
\def\pp{{\rm pp}}
\def\wic{\hbox{\rm WIC}}
\def\bigg{{\rm big}}



\setlength\arraycolsep{2pt}

\sloppypar

\setcounter{tocdepth}{2}
\fontsize{12}{13.5}
\addtocounter{page}{1}
\pagenumbering{arabic}

\setcounter{chapter}{9}

\title{A fusion of concepts}\label{10}
\titlerunning{Integral Forcing}

\author{Jin-Song von Storch} 
\authorrunning{von Storch}

\institute{
Jin-song von Storch, Max-Planck Institute for Meteorology}
\maketitle

\chaptermark{Integral Forcing} 
\index{von Storch, Jin-Song} \index{integral forcing}

In {\em Hasselmann Legacy
Stochastic Thinking in Climate Science}, 

\noindent Lin Lin and Hans von Storch (Editors),
to be published by Springer, 2026
\vspace{5mm}

\begin{abstract}{}

This essay fuses concepts and approaches used to describe fluctuating phenomena in climate systems and statistical mechanics, and explores new ideas essential for understanding such phenomena. Its starting points are the Langevin equation (LE) and the fluctuation–dissipation theorem (FDT). The former was introduced to climate research by Klaus Hasselmann through his stochastic climate models. While a version of the latter---formulated within the framework of linear response theory---has found wide application, the deeper origin of the relation between fluctuations and dissipation has remained inconclusive.
    This essay goes one step further by seeking the cause of the apparent randomness, rather than merely describing it as in the LE, and by directly linking a fluctuation–dissipation relation to the governing microscopic equations. It postulates that such a relation---also referred to as the integral fluctuation–dissipation relation (IFDR)---resides in integrals of the forcings that determine microscopic evolutions (individual trajectories) of the considered system.  
   The IFDR  ensures the emergence of well-defined macroscopic quantities, such as variance and spectra, quantities that characterize the system’s fluctuations, provided the system is in dynamical equilibrium with  constant external forcing. It is the dissipation embodied in IFDR that renders future states uncorrelated with initial conditions, thereby generating apparent randomness. Randomness is therefore not an unphysical artifact; on the contrary, it is a fundamental property of forced–dissipative systems in dynamical equilibrium. Fluctuation phenomena in such systems must be described by two principles---the governing microscopic equations and the IFDR. The two principles  are complementary but not reducible to one another.
    
\end{abstract}


\vspace{5mm}
Various concepts and approaches have been developed for  describing fluctuating phenomena in climate systems and in statistical mechanics. This essay focuses on fluctuations in forced-dissipative systems, which arise under time-independent external forcing, typically maintained by coupling to a reservoir. A defining feature of such systems is the stationarity of their fluctuations. In climate science, \IND{internal climate variability} provides the canonical example. In equilibrium statistical mechanics, representative cases include Brownian motion \citep{brown} and Johnson–Nyquist (thermal) noise \citep{johnson_1927}. In non-equilibrium statistical mechanics, which encompasses fluctuations in transient, driven, and aging systems, fluctuations in non-equilibrium steady states (NESS) play an analogous role. Deterministic dynamical systems offer yet another illustration, notably the fluctuations generated by the Lorenz model (\citeyear{lorenz_deterministic_1963}).
By placing fluctuations considered in statistical mechanics and those produced by deterministic chaotic dynamics within a single class of phenomena, we depart from the traditional classification based on their origins (stochastic versus deterministic) or thermodynamic attributes (zero versus positive entropy production). Instead, we identify time-independent external forcing as the common cause of these fluctuations, and we emphasize its  role, together with the inherent dissipation,  in establishing and sustaining their stationarity.

Concepts dealing with fluctuating phenomena were first developed in statistical mechanics. Klaus Hasselmann\index{Hasselmann, Klaus} was among the very few who introduced statistical mechanical concepts to climate science. Here, we will focus on two concepts:   the Langevin equation (LE)\index{Langevin equation} and  the fluctuation dissipation theorem (FDT)\index{fluctuation-dissipation theorem}.  

In Section \ref{s10.1}, we set the scene for studying fluctuating phenomena generated by a dissipative system subjected to time-independent external forcing. We do so by  introducing basic terms that are suitable for  describing all these   phenomena,  independent of whether they are found in the climate system or in  statistical mechanical systems.  In particular, we assume that all these systems are governed by autonomous differential equations in the form of Eq.(\ref{eq:x}). Moreover, because of the fixed external forcing, the system is uniquely described---both at microscopic level in terms of the system's solution and at macroscopic level in terms of time averages that converge with increasing averaging period and are consistent with the fixed external forcing. 

In Section \ref{s10.2}, we investigate properties common to these phenomena.  We argue that since the fluctuations of interest are generated by time-independent external forcings,  common properties can emerge because of the same forcing constraint, despite differences in processes involved in different systems. 

Section \ref{s10.3} is devoted to  the LE,  a concept developed for describing Brownian motion. We describe Langevin's original idea and how  this idea is transformed to Hasselmann's stochastic climate models, and why  the concept  is significant. We point out that in both the LE and  stochastic climate models, stochastic forcing is externally imposed. Introducing such a forcing does not  change the deterministic nature of  the system  in question. 

Section \ref{s10.4} extends the discussion beyond the LE by examining whether stochastic forcing is merely a modeling device or instead represents intrinsic randomness. If the latter holds, the system must be regarded as inherently random, which appears to conflict with the determinism of the governing differential forcings, such as $F({\bf x})$ in Eq.~(\ref{eq:x}), a deterministic function of the state vector ${\bf x}$.
This apparent tension is resolved by recognizing that randomness---understood here as the lack of temporal autocorrelation in the system’s solution---and determinism---defined by purely deterministic differential forcings like $F({\bf x})$---refer to distinct properties at different levels of description. Consequently, randomness in the solution does not contradict the determinism of the governing equations but instead constitutes a complementary characteristic of the system.

In Section \ref{s10.5}, various versions of the FDT are classified into two categories: one emphasizing the relation between fluctuation and dissipation under constant and reservoir-like external forcing, and the other highlighting the link between  the system’s response to external perturbations and the system's restoring property  found in the absence of perturbations.  This essay focuses on the former. In contrast to the LE where randomness is introduced externally, there exists a \IND{fluctuation-dissipation relation} (FDR) that  is inherent to systems in dynamical equilibrium with constant external forcing.  However, it remains unclear how such a FDR emerges from the underlying microscopic dynamics, typically taking the form of differential equations in time as given in Eq.(\ref{eq:x}). Does the FDR represents an intrinsic balance encoded in the differential forcing  $F({\bf x})$ in Eq.(\ref{eq:x})?

Sections \ref{s10.6} and \ref{s10.7} present a twofold extension of the classical FDT. First,  a direct connection is established between FDR and the underlying differential equations.  More specifically, we postulate that  an integral of a  differential forcing $F({\bf x})$ over time, which predicts for a given initial state the solution at a distant time,  constitutes a FDR. This FDR is encoded in time {\em integrals} of $F({\bf x})$, not in $F({\bf x})$ itself, and will be referred to as the \IND{integral fluctuation dissipation relation} (IFDR). Secondly,  the dissipation contained in time integrals of $F({\bf x})$---not that in $F({\bf x})$---is identified as being truly responsible for eliminating correlations between  states at sufficiently separated times,  and with that the emergence of the apparent randomness. This dissipation does not exist as a time rate of change.  The origin of this dissipation is further elaborated in  Section \ref{s10.8}.  

Section \ref{s10.9} emphasizes that both the dissipation embodied in integrals of $F({\bf x})$  and the inability of $F({\bf x})$ to predict  quantities defined as time averages that converge with increasing  averaging interval are manifestations of the same cause. Equilibrium fluctuations---conceptually existing over an infinite time axis---are governed by two complementary but distinct principles: the differential equations, which control the microscopic description, and the IFDR, which ensures a unique macroscopic description. 

The extension of the LE and that of the FDT, including the inability of $F({\bf x})$ to predict statistics of $x$ defined as converging time-averages, as well as the non-existence of IFDR as a time rate of change, are exemplified in terms of the Lorenz model (\citeyear{lorenz_deterministic_1963}).  Some concluding remarks are given in the final Section \ref{s10.10}. 

\section{Basic terms}\label{s10.1}

A fusion of different concepts is only possible when the phenomena they address can be described using a shared language. In this section, we define three basic terms, ``microscopic description'', ``dynamical equilibrium with a reservoir'', and ``macroscopic state''. Their definitions are aligned with, although  not exactly identical to, those used in statistical mechanics. We confine ourselves to classical systems.

\subsection{Microscopic description}\label{s10.1.1}

\IND{A  microscopic description} of a system relies on the principles that govern the system's temporal evolution. 
Consider a system under constant external forcing and described  by its state vector  ${\bf x}\in\mathbb{R}^N$ with $N>1$. These principles can be expressed in terms of a set of autonomous equations in  the form
\begin{equation}
    \frac{\mathrm{d}x}{\mathrm{d}t} = F({\bf x}(t)),
    \label{eq:x}
\end{equation}
each for a component $x$ of ${\bf x}$. $t$ indicates time. $F$---the differential forcing of $x$---is a deterministic, generally non-linear function of ${\bf x}$, which contains  parameters external to the system as well as processes that damp $x$.  For a different component $x'\neq x$, the respective $F$ is a different function of ${\bf x}$. 

A microscopic description \index{microscopic description} of a system is given by a solution of the set of  differential equations governing the system's evolution. Such a solution can generally only be obtained numerically by discretizing the time axis using an increment $\delta t$. The resulting discrete solution, obtained using a computer, will be denoted by  $\{x_i|i\in \mathbb{Z} \}$ with $\mathbb{Z}$ being the set of integers, or for short  $\{x_i \}$. The corresponding continuous solution can be obtained by letting $\delta t \rightarrow 0$.

For the climate, which has the atmosphere and the ocean as its major components,  the governing equations are  balance equations of momentum, energy, and matter. 
The most prominent external forcing  is the solar irradiance.  When representing the climate on a grid, as it is done in a climate model, the balance equations take the form of Eq.(\ref{eq:x}). The dimension $N$ of ${\bf x}$ equals the number of prognostic variables times the number of grid cells. $N$ is normally of the order of $10^7-10^8$, but may reach the order of $10^9 - 10^{10}$, depending on the available computer power.  

Fluctuating phenomena studied in statistical mechanics can be described by many-particle systems.  Within the realm of  classical physics, the evolution of such a system is governed by Newtonian mechanics. The state vector ${\bf x}$ consists of positions and momenta of particles. The dimension $N$ of ${\bf x}$ is  of the order of $10^{23}$ and hence much larger than that of a climate model. For an ideal system free of any dissipative process and not subjected to any external forcing,  the governing equations represent  the Hamiltonian dynamics with a constant Hamiltonian. Real physical systems  generally contain some kind of dissipation processes and have to be externally forced. 

Often, a microscopic description  is not possible without approximations. When describing the climate in terms of a  climate model, various types of approximations need to be made. These approximations range from neglecting components that can interact with the climate (e. g. ice sheet), to representing unresolved processes (e.g. small-scale turbulence)  using ad hoc  or empirical  relations that deviate from the exact governing principles. For  many-particle systems considered in statistical mechanics, we have furthermore the problem that the  dimension $N$ of  state vector ${\bf x}$ is incredibly large, rendering any attempt to solve the governing equations impractical.
Coarse-graining techniques, such as the Mori-Zwanzig projection operator method \citep{Mori1965,Zwanzig1961},  are  used. Within the Mori-Zwanzig formalism, the full set of governing equations is reduced to a set of equations for the relevant variables. These equations consist of two parts: one that describes the intrinsic dynamics of the relevant variables, and another that accounts for the influence of the irrelevant variables. The latter is captured through a memory term and a stochastic (random noise) term.

There is a fundamental difference in approximations used in climate research and statistical mechanics.  While  the unresolved processes are either completely ignored or  approximated using deterministic functions (parameterizations) in climate (at least traditionally), thereby keeping  the deterministic nature of  governing equations unchanged,  they are described statistically using random forcing in statistical mechanics. The choice to keep the governing equations of the climate purely deterministic reflects the desire of making  precise climate predictions.

A microscopic description can also be achieved through direct measurements, though the challenges involved differ from those faced when aiming at solving the governing equations. One of these challenges arises from the difficulty in controlling the external forcing factors. The real climate is influenced by  external forcing factors that vary with time, such as the variation in solar irradiance  due to solar activities,  changes in orbital parameters of the Earth, or an increase in atmospheric concentrations of greenhouse gases. As a result,  observed climate variations are not pure internal climate variations.  Other challenges are related to  the limited time span covered  by a measurement or limited frequencies resolved by a measuring instrument. Internal climate variations may not be fully observed since the length of observations is not long enough to record internal climate variations on  long time scales. Variations of particles considered in statistical mechanics may not be fully observed when the time scale of a particle is shorter than the reaction time of the observational eye. We return to this issue in Section \ref{s10.2}. 

\subsection{Dynamical equilibrium with a reservoir}\label{s10.1.2}

Dynamical equilibrium with a reservoir (or reservoirs), \index{dynamical equilibrium with a reservoir}  for short dynamical equilibrium,  includes thermal equilibrium and non-equilibrium steady state---NESS. The former is used in equilibrium statistical mechanics. A system is said to be in thermal equilibrium with a heat bath when it has the same temperature as the heat bath. The latter is used in non-equilibrium statistical mechanics. A system is said to be in a NESS when it is driven out of thermodynamical equilibrium, but is in a time-independent macroscopic state despite ongoing microscopic fluctuations. As a consequence of external forcing, there are ``currents'' of energy, momentum, particles, etc., so that the system under consideration is open to exchanges of these  quantities with its environment. 

The term ``dynamical equilibrium with a reservoir'' emphasizes a fundamental trait shared by systems in thermal equilibrium and in NESS: both are continuously driven by a constant and reservoir-like external forcing, whether arising from  a heat bath or via any other mechanism. The term ``reservoir'' underscores the infinite capacity, with which the external forcing drives the system. 
For the climate, the Sun plays the role of a reservoir. The divergence of the net radiative fluxes, that result from radiative interactions of the Sun with the planet Earth, provide the energy for a broad spectrum of atmospheric and oceanic motions. For Brownian particles immersed in a glass of water and for electrons in a resistor,   the surrounding air (at a nonzero room temperature) acts as a reservoir that provides energy for water molecules and electrons. Dynamical equilibrium with a reservoir can only be achieved by a dissipative system. 

In reality, an ideal reservoir does not exist. Instead, the term ``reservoir'' refers to the  situation, where the energy required for the system's state vector ${\bf x}$  to vary is negligibly small relative to the energy available from the  reservoir. In case of climate, the energy needed for atmospheric and oceanic variations is negligible relative to the energy available from the Sun. In case of thermal noise and Brownian motion, the energy needed for   electrons or water molecules to move is negligible relative to the energy available from the surrounding air. 

We note that for systems in dynamical equilibrium with a reservoir, the time needed for the system to reach this equilibrium varies from system to system.  While Brownian particles immersed in water reach an equilibrium with the surrounding air  almost instantaneously, a climate simulated by a climate model can only reach an equilibrium   after the model has been run with  the same constant external forcings  over a sufficiently long time period. Normally, a climate simulation forced with constant external forcing is  referred to as a control simulation.  It is a well-accepted practice that a control simulation has to be preceded by a  spin-up simulation, during which the model drifts from an initial state that is not consistent with,  to a state that is fully consistent with, the  external forcing. The length of a spin-up simulation is thought to be   at least a few decades,  but can be longer than hundreds or more years. 

We also note that radiative equilibrium in a climate system does not necessarily entail dynamical equilibrium. Radiative equilibrium primarily constrains the global mean temperature, but  places no direct constraints on other climate variables. More importantly, it does not restrict the presence or magnitude of fluctuations in climate variables.

\subsection{Macroscopic state}\label{s10.1.3}

A macroscopic state \index{macroscopic state} is described by macroscopic quantities. In statistical mechanics, macroscopic quantities are typically defined as ensemble averages. Time averages along trajectories (solutions) are generally not  considered, as the trajectories themselves are often unknown or too difficult to compute. While this ensemble-based approach is convenient, it also introduces certain complications.

An important  complication is the need for {\em different} ensembles---defined by different probability functions, in order to account for different physical  conditions. This is why we distinguish, for example, between microcanonical and canonical ensembles in equilibrium statistical mechanics. Similarly, in non-equilibrium statistical mechanics, studying chaotic dynamical systems in a NESS requires using the SRB (Sinai–Ruelle–Bowen) measure, which defines the appropriate probability distribution for describing the ensemble in the long-time limit. 

However, it is generally challenging to derive the distribution function that defines an ensemble. One possibility is to obtain it directly from the underlying differential equations. This immediately raises the difficulty of reconciling the purely deterministic nature of these equations with the stochastic character implied by a distribution function.
Thus,  the ensemble-based approach  sidesteps the deeper question of whether the system itself is truly stochastic. The usual justifications for a probabilistic treatment---such as the inability to track all degrees of freedom, to specify initial conditions precisely, or to fully resolve the dynamics---emphasize factors that are {\em external} to the system. These arguments only further underscore the deterministic nature of the system, thereby rendering a probabilistic consideration conceptually unsatisfying.

When the system's solution is known and its fluctuations are stationary and incessant, the above complications can be circumvented by employing time averages along solution trajectories. Being a deterministic functional of the solution,  such an average does not depend on any probability measure.
There is no need to adopt a probabilistic perspective for defining macroscopic quantities. Stationarity ensures that, along a solution, time averages of the state vector ${\bf x}$, or functions thereof, converge as the averaging time increases. These time-averages are hence definite---not containing any uncertainty, and will be used to define  macroscopic quantities.

Examples of \IND{macroscopic quantities} are the mean $\mu$ and the variance $\sigma^2$ of a component $x$ of ${\bf x}$. For a discrete solution of $x$,   these quantities are defined as 
\begin{subequations} \label{eq:equilibrium}
\begin{align}
  \mu =& \lim_{n\rightarrow \infty} \frac{1}{n}\sum_{i=1}^n x_i \equiv \lim_{n\rightarrow \infty} \overline{x}^n , \label{eq:equilibrium_mean} \\
 \sigma^2=  &  \lim_{n\rightarrow \infty} \frac{1}{n}\sum_{i=1}^n \big (x_i-\overline{x}^n \big )^2  .  \label{eq:equilibrium_var}
\end{align}   
\end{subequations}

\section{Equilibrium fluctuations}\label{s10.2}

Equilibrium fluctuations \index{equilibrium fluctuations} are fluctuations in a system that is in dynamical equilibrium with a reservoir. Internal climate variations and fluctuations related to Brownian motion and thermal noise  are all equilibrium fluctuations. In Subsection \ref{s10.2.1}, we explore the three key properties of equilibrium fluctuations. Among them, the third---concerning the portion of the fluctuation variance at zero and near-zero frequencies---is often overlooked. Yet this property is crucial for understanding equilibrium fluctuations as a phenomenon that spans all time scales, from infinitesimally short to infinitely long. The fact that an infinite time scale is not merely a mathematical construct but a physical necessity follows directly from the involvement of a reservoir. Subsection \ref{s10.2.2} examines this third property in greater detail.

\subsection{Properties of equilibrium fluctuations}\label{s10.2.1}

Equilibrium fluctuations have the following  properties in common. First,  being in contact with a reservoir, equilibrium fluctuations are stationary and incessant. This property has been used in Section \ref{s10.1.3} for  defining macroscopic quantities. Every time when we look through a microscope (in a room with a nonzero temperature), we see similar zigzag motions of Brownian particles. Every time when we turn on a  vacuum tube radio (in a room with a nonzero temperature), we are confronted with similar thermal noise. When simulating a climate  using a climate model subjected to constant external conditions,  the simulated climate  will (after a proper spin-up) vary  continuously in the same fashion as long as the simulation goes. 

Secondly, equilibrium fluctuations appear to be random. Here, the term \say{random} does not refer to the set of governing equations, but rather to their solutions.   A solution of $x$ is said to be random, if for a sufficiently large value of $\tau$, the solution at the $k$-th time step is uncorrelated with the solution at the $k+\tau$-th time step. In other words, the  auto-covariance function $\gamma_\tau$ of $x$ effectively vanishes for sufficiently large values of  $\tau$. Here,  $\gamma_\tau$  is defined in terms of the time average in  long-time limit, similar to the mean and the variance in Eq.(\ref{eq:equilibrium}), 
\begin{equation}
    \gamma_\tau= \lim_{n\rightarrow \infty} \frac{1}{n}\sum_{i=1}^n \big (x_i-\overline{x}^n \big ) \big (x_{i+\tau}-\overline{x}^n \big) .
    \label{eq:acf}
\end{equation}   
Even though  difficult to verify, our experience tells us that Brownian motion and thermal noise, as well as variations of meteorological variables,  are random in the  sense of  lacking auto-correlations at sufficiently  large time lags.  

Fluctuations of $x$ can also be described by   the spectrum  of $x$,  $\Gamma^x(\omega)$. We consider only spectrum defined as a function of  positive frequency $\omega$. The third property says that $\Gamma^x(\omega)$ is continuous and has a white and nonzero low-frequency extension.
Here, \say{low-frequency} refers to a frequency range $[0,\omega_0]$, including  the lowest possible frequency $\omega=0$. The continuity  implies that in the limit $\omega\rightarrow 0$,   the low-frequency extension equals \IND{$\Gamma^x(0)$}.   Two spectra with the asserted shape are  sketched in Fig.\ref{fig:sp_eye_dyn}. 

\begin{figure}[ht]
\includegraphics[width=0.95\textwidth]{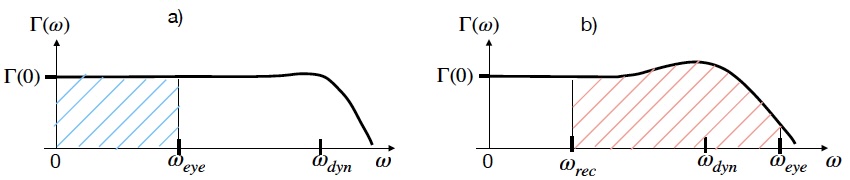} 
\caption{Sketch of spectrum $\Gamma^x(\omega)$  of a component of  a  system in dynamical equilibrium with a reservoir. The frequency axis is logarithmic, emphasizing the low frequencies. $1/\omega_{eye}$, $1/\omega_{dyn}$ and $1/\omega_{rec}$ indicate respectively, the response timescale of the observational eye, the major timescale of the underlying dynamics, and the length of the record  used to derive the spectrum. The blue hatched area indicates the part of the total variance found, when the observational eye is unable to resolve the high-frequency part of the spectrum. The red hatched area indicates the  part of the total variance found, when the record is too short to cover the low-frequency part of the spectrum. }
\label{fig:sp_eye_dyn}
\end{figure}

\subsection{More about \IND{$\Gamma^x(0)$}}\label{s10.2.2}

The third property implies that  a portion of the total variance $\sigma^2$  of $x$ is associated with fluctuations of $x$ on arbitrarily long time scales. This may feel conceptually unsettling. Is $\Gamma^x(0)$ physically real and observable?  

The answer  is yes,  even though it can be difficult to identify $\Gamma(0)$ in practice. For thermal noise, a white spectrum is generally assumed, except at extremely high frequencies.  
The assumption, which works well for signal processing used in communications and electronics, implies a nonzero $\Gamma^x(0)$. For Brownian motion, it was found that the typical observed velocity of a Brownian particle is smaller than the typical velocity predicted by the equipartition theorem. According  to \cite{MacDonald},  this is because the equipartition theorem predicts the total variance of the velocity, while an observer can only observe a part of this total variance,  due to  the limited response time of the eye.  The situation is sketched in Fig.\ref{fig:sp_eye_dyn}a). MacDonald's explanation underscores that Brownian velocity fluctuates over   a frequency range extending to an unimaginably long  time scale (from the point of view of a Brownian particle), thereby pointing to the existence of a nonzero $\Gamma(0)$.  

For a climate in equilibrium with the Sun (and other external forcing factors), a typical spectrum of equilibrium climate variations is sketched in Fig.\ref{fig:sp_eye_dyn}b). Even though  the response time of the observational eye $1/\omega_{eye}$ is generally no longer an issue,  it is still, or even more, difficult to derive the full  spectrum, and with that $\Gamma(0)$, from measurements. The difficulty is now caused by the finite length of observational records,  $1/\omega_{rec}$, and the fact that the real climate is  influenced by  time-varying external forcings.  Spectra with white low-frequency extension can only be observed for climate variables, whose  variations  are predominantly generated by internal dynamics that operate under constant external forcings, and are not  significantly influenced by any time-varying external forcing. 

\begin{figure}[htb!]
\includegraphics[width=\textwidth]{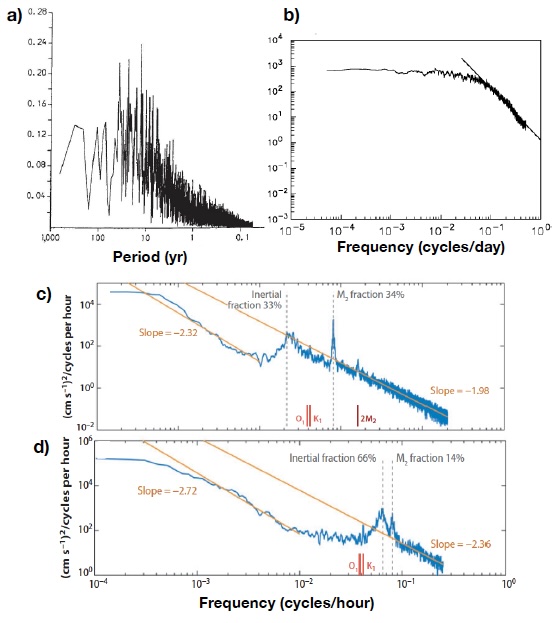} 
\caption{ Spectra of a) a prognostic variable in a simulation performed with an atmospheric model (adopted from \cite{jamesjames}), b) surface pressure measured at the station Potsdam (adopted from \cite{JSvonstorch:01}), c) \& d) current kinetic energy from instrumental records in the North Atlantic at 500 m and in the South Pacific at 1000 m (adopted from \cite{FerrariWunsch}}
\label{fig:JJDFW}
\end{figure}

Figure \ref{fig:JJDFW} displays a few climate spectra. 
The spectrum depicted in Fig.\ref{fig:JJDFW}a) is obtained from a simulation performed with an atmospheric model subjected to a constant external forcing, namely the solstice forcing \citep{jamesjames}. Even though the major internal dynamics---the baroclinic instability---operates on timescales of days, the spectrum of a component of the state vector ${\bf x}$ extends to timescales longer than years, decades, and even centuries, albeit with poor spectral quality. 
The spectra in Fig.\ref{fig:JJDFW}b)-d) are obtained from direct measurements.  All of them seem to have a low-frequency white extension, even though the records may not be long enough, especially for the current spectra, to allow a definitive conclusion. 

A more decisive conclusion can be drawn from  solutions of the Lorenz 1963  model---a prototype of dynamical systems in NESS. 
Due to its low dimension,  we can easily produce sufficiently long and sufficiently many solutions to derive Lorenz spectra that are sufficiently accurate and  extend to smaller and smaller frequencies. Fig.\ref{fig:lorz_sp} shows that for all three Lorenz components, a white nonzero low-frequency extension emerges and extends to increasingly lower frequencies as the solution length increases.  
Once emerged, the white low-frequency extension stays at the same level, while the spectral level at the highest frequencies drops with increasing solution length.
The drop is understandable, since the total variance, defined as the  time average in long-time limit according to Eq.(\ref{eq:equilibrium}b),  converges fast with increasing solution length and is essentially the same across the four cases shown in Fig.\ref{fig:lorz_sp}. 
As the solution length increases, progressively lower frequencies of the spectrum are resolved, leading to a larger contribution of the low-frequency part to the total variance. To preserve the total variance, the contribution from the high-frequency part must therefore diminish. Consequently, for each of the three Lorenz components, a well-defined finite nonzero $\Gamma^x(0)$  can be obtained {\em asymptotically} by increasing solution length.

\begin{figure}[hbt!]
\includegraphics[width=0.9\textwidth]{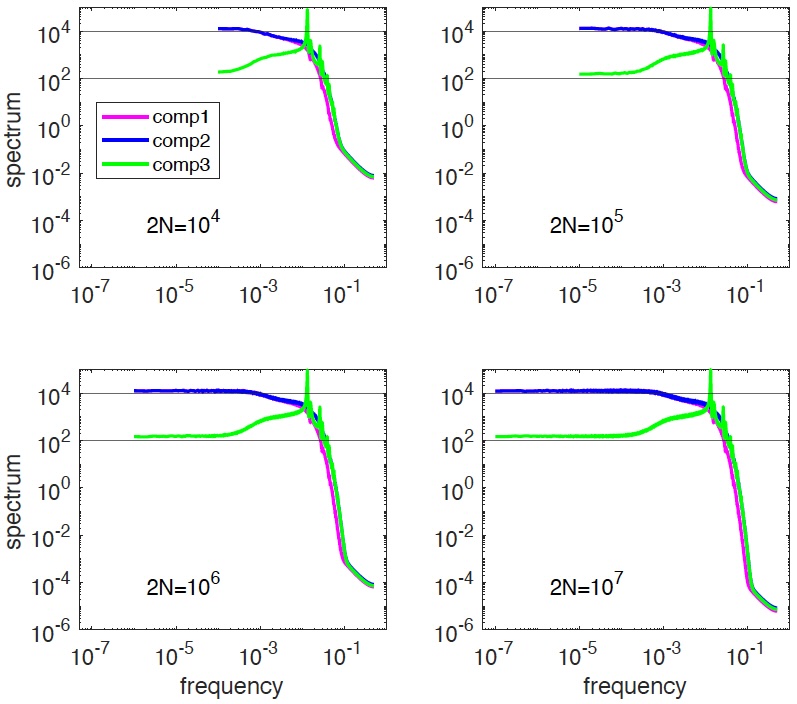} 
\caption{Spectra of the three Lorenz components (magenta, blue, green), estimated as ensemble-averaged periodograms using ensembles of equilibrium solutions of length $2N+1$ with $2N=10^4,~10^5,~10^6, ~10^7$ and an ensemble size $m=1000$.   Each equilibrium solution  is obtained by integrating the Lorenz 1963  model  from a quasi-equilibrated state, using  a time step of 0.01. Even though the spectrum is defined as Fourier transform of auto-covariance function, the ensemble-averaged periodogram  is used, since a)  the latter converges to the former in the limit $N\rightarrow \infty$ and $m\rightarrow \infty$ (see Appendix B in \cite{JvS:22}), and b) the convergence of ensemble-averaged periodogram is easier to achieve than the convergence of auto-covariance function  at large $\tau$. }
\label{fig:lorz_sp}
\end{figure}

The three properties of equilibrium fluctuations are interlinked.  More specifically, the first two properties serve as prerequisites for the existence of $\Gamma^x(0)$.  The first property makes the spectral value at  frequency zero physically meaningful, since only solutions that vary incessantly  can possess variations on infinite timescales. The second property, which  implies that the auto-correlation function must be effectively zero for sufficiently large $\tau$, ensures the existence of  spectrum  as  Fourier  transform of auto-covariance function $\gamma_\tau$, or in case of a real-valued $\{x_i\}$, the cosine Fourier transform of $\gamma_\tau$:
\begin{equation}
    \Gamma^x(\omega)=\sum_{\tau=-\infty}^{\infty} \gamma_\tau \cos(2\pi\tau\omega).
    \label{eq:cosFT}
\end{equation}
This $\Gamma^x(\omega)$ must be continuous, since the cosine function is continuous.  This spectrum must also be white at near-zero frequencies, since  the cosine function is flat at the origin.  In the limit $\omega\rightarrow 0$, the level of the low-frequency extension must approach
\begin{equation}\label{eq:gamma0}
 \Gamma^x(0)=  \sum_{\tau=-\infty}^{\infty} \gamma_\tau.
\end{equation}
$\Gamma^x(0)$ is expected to be nonzero, since $\gamma_\tau$  has its maximum at $\tau=0$.  $\Gamma^x(0)$  is  expected to be  finite, despite the summation over an infinite number of summands, since  $|\gamma_\tau|$ decays with increasing $\tau$.   We hence conclude that  a  finite and non-zero $\Gamma^x(0)$ is a manifestation of the randomness in  solution of $x$. 

There is another way to see the link between $\Gamma^x(0)$ and our definition of randomness (the second property of equilibrium fluctuations).  The vanishing auto-covariance at large lags implies  that a random solution cannot be periodic in time. The spectrum of a periodic solution is markedly different from that of a non-periodic solution. 
A periodic solution with period $P$ can be represented by a sum of sine and cosine functions with well-defined Fourier coefficients \citep{priestley1981}. The associated spectrum is {\em discrete} and consists of spectral lines at frequencies $\omega_k=k/P$ with $k=0,1,\cdots$, each given by the square of the respective Fourier coefficients. However, only those with $k\geq 1$ represent the {\em variance} of  an oscillation with period of $P/k$.  The one with $k=0$ represents the square of the {\em time mean}. It does not represent the  variance of an oscillation with an infinite period. A periodic function with an infinite period  is ill-posed, since it cannot cycle and repeat itself in the classical sense.  In other words, $\Gamma(0)$ does not exists as a quantity describing the variance for a periodic solution.

Conversely, the spectrum of a random solution, characterized by vanishing auto-correlations at sufficiently large lags,  cannot be discrete. For a random discrete  solution $\{x_i\}$ having an infinite length,  its Fourier transform---an operation applied on an  infinite series---does not exist, since this series $\{x_i\}$ is not absolutely summable  \citep{priestley1981}. For a finite piece of a random solution, we may obtain Fourier coefficients  by applying the Fourier analysis to the piece, and use these   coefficients  to define a discrete spectrum. However,  we find that  Fourier coefficients at the same frequencies are different for different pieces,  even though the pieces stem from the same solution.   Thus, the spectrum of a random solution cannot be described by discrete spectral lines. It has to be defined as the Fourier transform of auto-covariance function, as given in Eq.(\ref{eq:cosFT}), which gives rise to a continuous    spectrum with a nonzero finite $\Gamma^x(0)$. 
Thus, for a random solution, a  nonzero finite $\Gamma^x(0)$ is both mathematically justified and physically meaningful. 

\section{Langevin's equation and Hasselmann's stochastic climate models}\label{s10.3}

At the time when Hasselmann introduced his \IND{stochastic climate models},  climate variations were understood to be generated by two types of mechanisms. One  relies on  variations in the external forcing. Climate variations are thought to be responses to temporal variations in the external forcing,  such as variations in solar irradiance due to  changes in Earth's orbital parameters,  volcanic eruptions that increase the turbidity of the atmosphere and with that reduce the solar radiation arriving at the  Earth's surface, last but not least changes in the atmospheric concentration of green house gases. 
The other type of mechanisms emphasizes internal atmosphere-ocean-cryosphere-land feedbacks, especially the positive ones. A positive feedback can amplify the response of the climate to changes in its external forcing. If sufficiently strong, a positive feedback can produce unstable spontaneous transitions from one climate state to another. Both types of mechanisms arise from a purely deterministic thinking. 

Hasselmann suggested a third type of mechanism, inspired by two statistical mechanical treatments of Brownian motion. One utilizes the Fokker-Planck equation that describes the evolution of the probability distribution of a variable related to Brownian motion. The other takes the form of a Langevin equation that describes the typical behavior of a Brownian particle based on Newtonian mechanics. The two treatments are mathematically equivalent. We should concentrate on the latter, as it is  more intuitive than the former.  

Langevin's equation \citep{langevin,lemons} \index{Langevin equation}  is based on Newton’s second law that a mass  times its acceleration equals the force acting on the mass. For a  Brownian particle  immersed in water and having mass $m$ and velocity $v$,  this equation can be written as
\begin{equation}
    m\frac{\mathrm{d}v}{\mathrm{d}t} = -\lambda v+\zeta.
    \label{eq:le}
\end{equation}
The right hand side of Eq.(\ref{eq:le})  represents two forces acting on the Brownian particle. The first one,  $-\lambda v$, referred by Langevin as a \say{viscous resistance}, is a frictional force. The positive constant $\alpha$ determines the strength of this friction and may be  obtained from Stokes’ formula.  The second one,  $\zeta$, is referred by Langevin as a \say{complementary force} that is  \say{indefinitely positive and negative}. It results from the impact of the particle with water molecules. Without a detailed description, this impact is represented by   a  delta-correlated white noise. 

The Langevin equation is an approximation of the set of equations governing the evolution of  Brownian motion. It has been thought that the approximation works excellently  in the Markovian limit in which the correlation time of variables describing  water molecules  is negligibly short relative to that of variables describing  the Brownian particle.  The latter  is manifested in the timescale of the frictional force.  In case when the Markovian limit is not met, the Langevin equation can be generalized (see e.g. \cite{Balakrishnan}). In a generalized Langevin equation, the frictional force is no longer proportional to instantaneous velocity, but has to be described in terms of  a memory kernel. The noise term is no longer described by  a white noise, but by a colored one. 

In order to use Brownian motion as an analogy, Hasselmann had to identify \say{Brownian particles} and \say{water molecules} in the climate system, and with that   a Markovian limit suitable for the climate system.   He did so by assuming that the climate system consists of  slow and fast components with distinctly different timescales. The fast components were  identified with the  “weather” variables, whereas the slow components were thought to be associated with variables such as the sea surface temperature, sea ice, and land vegetation.  Hasselmann then asserted  that variations in  slow climate components are driven by weather, which acts as a stochastic forcing, very much like $\zeta$ in Langevin's equation.  For a slow component $x$ governed by an equation in form of Eq.(\ref{eq:x}), Hasselmann's stochastic climate model \index{Stochastic Climate Model} takes the form
\begin{equation}
    \frac{\mathrm{d}x}{\mathrm{d}t} = \widetilde{F}+\zeta.
    \label{eq:scm}
\end{equation}
$\widetilde{F}$ represents the dynamics of the slow component $x$ averaged over a time period shorter than the timescale of $x$ but longer than the timescale of weather. $\zeta$ is a stochastic forcing arising from weather. The similarity between Eq.(\ref{eq:scm}) and Eq.(\ref{eq:le}) is obvious.  

As Langevin’s 1908 paper inspired new mathematics (stochastic differential equations) and new physics (stochastic physics), Hasselman's  1976 paper  has inspired  stochastic considerations in climate research. However, what sometimes gets lost in these considerations is that the mechanism represented by Hasselmann's stochastic climate model relies on the joined action of {\em two} ingredients, not just that of the stochastic forcing $\zeta$.
To see this, we note  that in the absence of $\widetilde{F}$, the solution to $\frac{\mathrm{d}x}{\mathrm{d}t} = \zeta$ is a random walk,  whose variance increases with increasing time. Thus, in order to  ensure an equilibrium climate, $\widetilde{F}$ must contain, in Hasselmann's language,  \say{negative feedbacks}. The need for \say{negative feedbacks}  is in striking contrast to the appreciation for positive feedbacks that prevailed back in the 1970s. 
The need complies with the fact that Hasselmann's stochastic climate model targets at internal variations in an equilibrium climate.  

There is one difference between Hasselmann's stochastic climate model and  Langevin's equation. While  the “viscous resistance” $-\lambda v$ is linked to the stochastic forcing $\zeta$ in  Langevin's equation,  as both  originate from collisions of a Brownian particle with water molecules, there is no link between "negative feedbacks" in $\widetilde{F}$ and the stochastic forcing $\zeta$ arising from weather variables in Hasselmann's stochastic climate models. 

We point out that in both the LE and stochastic climate models, stochastic forcing is {\em externally} imposed for practical reasons.
In the case of Brownian motion, the system involves an overwhelmingly large number of degrees of freedom. A stochastic forcing is  introduced to represent the effect of interactions among these degrees of freedom, without explicitly resolving them. 
In a stochastic climate model, the system under consideration is typically decomposed into a slow and a fast subsystem. For the slow subsystem---the focus of a stochastic climate model, the separation inevitably leads to a closure problem. A stochastic term is added to represent the effect of the fast processes on the slow dynamics. 

\section{An apparent contradiction and its solution}\label{s10.4}

According to section \ref{s10.2}, equilibrium fluctuations {\em are} inherently random, which seems to contradict the deterministic nature of the forcing term $F({\bf x})$ in  Eq.(\ref{eq:x}). The contradiction is culminated in $\Gamma^x(0)$.

Because of Eq.(\ref{eq:x}), the spectrum of $x$, $\Gamma^x(\omega)$, and the spectrum of  $F$, $\Gamma^F(\omega)$, are related to each other via
\begin{equation}
    (2\pi\omega)^2 \Gamma^x(\omega)=\Gamma^F(\omega).
    \label{eq:sp_xf}
\end{equation}
This is a  forcing-response relation. Given the variance of $F$ around frequency $\omega$, $\Gamma^F(\omega)\mathrm{d}\omega$, we obtain  the variance  of $x$ around the same frequency, $\Gamma^x(\omega)\mathrm{d}\omega$, weighted by the gain function $(2\pi\omega)^2$, which arises from the differential operator $\frac{\mathrm{d}}{\mathrm{d}t}$.

For equilibrium fluctuations of $x$ characterized by a finite and nonzero  $\Gamma^x(0)$, Eq.(\ref{eq:sp_xf}) reduces for $\omega=0$ to 
\begin{equation}
\Gamma^F(0)=0.
\label{eq:sp_F_0}
\end{equation}
The nonzero variance $\Gamma^x(0)\mathrm{d}\omega$ cannot be determined  by $\Gamma^F(0)=0$. It can also not be determined by $\Gamma^F(\omega)$ at any nonzero frequency, since  Eq.(\ref{eq:sp_xf})  holds frequency-wise.  We must conclude that $\Gamma^x(0)\mathrm{d}\omega$, and hence also the variance of $x$, cannot be determined by $F$  (for more detailed discussion see \cite{JvS:22}). The difficulty in identifying the spectral relation between $x$ and $F$ due to  vanishing $\Gamma^F(\omega)$ with decreasing frequency was first recognized and discussed in the context of the atmospheric axial angular momentum simulated by a climate model \citep{vonstorch1999red,vonstorch1999b}.

The contradiction, which exposes the inability of $F$ in determining the variance of $x$, can be explained by adopting  the definition of macroscopic quantities given in section \ref{s10.1.3}. According to this definition, an auto-covariance function---no matter whether it is $\gamma^x_\tau$ that defines $\Gamma^x(\omega)$ or $\gamma^F_\tau$ that defines $\Gamma^F(\omega)$---equals the converge of a time-average obtained when the averaging period approaches infinity. Such a convergence does not comply with $F$---a time rate, defined without any time integration. In this sense,  the randomness in a solution of $x$ and the determinism in the differential forcing  $F$  of $x$ refer to two different aspects of equilibrium fluctuations of $x$. Randomness refers to a property that emerges in its purest form {\em after} having integrated Eq.(\ref{eq:x}) over a sufficiently long time interval.  Determinism on the other hand refers to the determinism in $F$  {\em before} Eq.(\ref{eq:x}) is integrated over time. Being referring to two different things,  the contradiction is  not a real one and dissolves by itself.  


\section{Fluctuation-dissipation theorem}\label{s10.5}


The \IND{fluctuation-dissipation theorem} (FDT) has come in different forms.  Early forms  include Einstein's consideration addressing diffusion and  Nyquist's explanation for thermal noise\index{Nyquist's fluctuation-dissipation theorem}.  The former relates the diffusion coefficient (describing fluctuations in position) to the mobility  \citep{Einstein1905}. The latter relates the power spectral density of voltage (due to fluctuations of electrons in a resistor)  to resistors' resistance  \citep{nyquist}. \cite{CallenWelton}  proved that Nyquist's relation can be extended to a general class of linear dissipative quantum systems.   \cite{kubo_1957,kubo_1966}
extended the FDT within the framework of linear response theory. His key result links the response of a system to an external perturbation with the auto-covariance function obtained in the absence of the perturbation. Inspired by Kubo’s work, the FDT has been 
applied to a wide range of systems.
One important class of systems is the chaotic deterministic dynamical systems in NESS (recall the broad definition given  in the opening paragraph of this essay).
For those, Ruelle specifically developed his response theory  to describe how small perturbations affect their behavior \citep{Ruelle1998,Ruelle2009}. 
More recent developments, reviewed for instance by \cite{Marconi2008} and \cite{Baldovin2022}, focus on constructing theoretical frameworks to address responses in out-of-equilibrium systems, where the classical FDT no longer straightforwardly applies.

Intuitively, the various forms of FDT can be grouped into two fundamental classes.
The first one pertains to systems in thermal equilibrium and expresses a direct relation between dissipation and variance of fluctuations. This class of the FDT was addressed by Nyquist in the context of electrical noise and later formalized and generalized by Callen and Welton for quantum systems.
The second class applies to systems that are subjected to small external perturbations, whether near or far from equilibrium. It relates the linear response of the system to perturbations with unperturbed correlation functions or relaxation properties of the system. Such a relation arises because dissipation, which can be considered as a restoring force operating already in the unperturbed system,  affects  the system’s response: the stronger the dissipation, the stronger the inherent restoring force, the weaker the response.  Since the work by Kubo, the   FDT  has been primarily understood through the lens of response theory. 

Einstein’s relation occupies a somewhat distinct place in the landscape of FDT. It does not fit cleanly into the first class, because it connects the diffusion coefficient (a measure of fluctuations) to the mobility, which is (in case of Brownian motion) inversely related to dissipation.  Nor does it fall squarely into the second class, because it links mobility---interpretable as the steady-state response to a constant force---to diffusion coefficient, rather than to a correlation function.
The distinction arises because diffusion results from fluctuations in position, which accumulate over time so that their variance grows over time, unlike fluctuations in velocity. This cumulative nature gives position-based fluctuations a qualitatively different character, making Einstein’s relation structurally different from FDTs based on velocity auto-correlations.

Kubo’s fluctuation-dissipation theorem (FDT) was introduced to the climate science community by \cite{Leith1975}. One of its generalizations---Ruelle’s response theory---has  been extended and applied to climate systems by \cite{lucarini2017}. Despite their theoretical importance, these response-based approaches are not the standard for predicting climate responses to external forcings, such as an increase in  greenhouse gas concentrations in the atmosphere. Instead, climate projections continue to rely primarily on direct numerical simulations, as reflected in IPCC assessment reports (see e.g. \cite{IPCC6_Chap4_Future_global_climate}). This preference is due not only to the growing maturity of climate models, but also to the fact that direct simulations do not require the weakness of external perturbations---an assumption that underlies the linear response theory.

Unlike the response-based FDT, the fluctuation-dissipation relation developed by Nyquist and later extended by Callen and Welton has not found application  in the context of climate science.  That Nyquist's explanation considers only a highly special case is obvious. To see that the consideration of Callen and Welton is likewise  not directly applicable  to the climate system, we consider the energy levels.  For thermodynamic systems considered by Callen and Welton,  the relevant energy level is on the order of $k_BT$.  With the Boltzmann constant $k_B$ of about $1.4\times 10^{-23}$ J/K and room temperature  $T \approx 300$ K, this yields a characteristic energy of roughly $4\times 10^{-21}$ J. In contrast, the kinetic energy of geophysical flows is many orders of magnitude larger. For example, air with density $\rho=1.2$ kg/m$^3$ and velocity $v=1 m/s$ in $1$ m$^3$ volume carries about 0.6 J of kinetic energy. Sea water at 10 cm/s in the same volume carries about 5 J. This vast energy gap reflects a fundamental difference: thermodynamic systems require quantum descriptions due to energy quantization---as it is done by Callen and Welton, while climate systems are governed by classical, Newtonian dynamics. 

Even though Nyquist’s explanation of thermal noise and the broader consideration of Callen and Welton cannot be directly applied to the climate system, the  central idea---that fluctuation must be linked to dissipation to maintain a dynamical equilibrium---is general. 
Below, we adopt this general idea and assume that for a system in dynamical equilibrium with a reservoir, dissipative behavior is intrinsically linked to fluctuations:   Anything that generates fluctuations must also damps them;  anything that damps fluctuations must also generates them. 

\section{Integral fluctuation-dissipation relation (IFDR)}\label{s10.6}

To make the idea of Callen and Welton concrete for equilibrium fluctuations in classical systems, 
we begin with the governing equations in the form of Eq.(\ref{eq:x}). 
Given that randomness is defined by the lack of correlation between $x_i$ and $x_{i+\tau}$---the solution at two different times,  we consider the definite integral of $F$ that determines  $x_{i+\tau}$ for  given $x_i$. 
When discretizing time using increment $\delta t$ and denoting $F$ at $j$-th time step by $F_j\equiv F(\bf x_j)$,  an integral of $F$ over a time interval of length $\tau \delta t$ can be written as
\begin{equation}\label{eq:g_def}
G_{\tau,i} = 
    \sum_{j=i}^{i+\tau-1}F_j \delta  t ~~~\text{for}~\tau \in \mathbb{Z}_+,
\end{equation}
where $\mathbb{Z}_+$ indicates the set of positive integer. 
Hereafter, $G_{\tau,i}$ is referred  to as an \IND{integral forcing}.
The first subscript of $G_{\tau,i}$  indicates the length of the time interval, over which $F$ is integrated. The second subscript is a running index indicating the time at which the integration starts. It holds
\begin{equation}
    x_{i+\tau}=x_i+G_{\tau,i}.
    \label{eq:x_int}
\end{equation}
Because of Eq.(\ref{eq:x_int}), whether or not  $x_i$ is correlated with $x_{i+\tau}$  depends solely on the properties of the respective integral forcing $G_{\tau,i}$. 

We postulate that for any given nonzero value of $\tau$, $G_{\tau,i}$  consists of and only of, apart from a constant $c_\tau$, a dissipation   $d_\tau x_i$ and  a fluctuating forcing $f_{\tau,i}$,
\begin{equation}
    G_{\tau,i} = c_\tau+ d_\tau x_{i} + f_{\tau,i}, ~~~\text{for}~\tau \in \mathbb{Z}_+.
    \label{eq:g_df}
\end{equation}
Eq.(\ref{eq:g_df}) constitutes the \IND{integral fluctuation-dissipation relation}.
$d_\tau$ is negative and takes values in the interval $(-2,0)$.   The strongest dissipation, defined  by complete elimination of $x_i$, is reached when $d_\tau=-1$.  There is no further constraint on $f_{\tau,i}$, apart from being stationary and having a non-zero variance. $d_\tau$ and $f_{\tau,i}$ describe two different effects of the integrated $F$. 

For a fixed value of $\tau$, $c_\tau$, $d_\tau$ and $f_{\tau,i}$ can be obtained by regressing $G_{\tau,i}$ against $x_i$. $c_\tau$ is then the intercept, $d_\tau$ the regression slope, and $f_{\tau,i}$ the residual. 
For a system in dynamical equilibrium,  $d_\tau$  and $c_\tau$  obtained from regression using   $n$ pair of $(x_{i\tau}, G_{\tau,i\tau})$ converge with increasing $n$, in analog to the convergence in Eq.(\ref{eq:equilibrium}) and Eq.(\ref{eq:acf}). $d_\tau$ can be shown to be linked to the auto-correlation function $\rho_\tau$ of $x$ via
\begin{equation}\label{eq:d_rho}
    1+d_\tau =\rho_\tau.
\end{equation}

The integral forcing $G_{\tau,i}$ in Eq. (\ref{eq:g_df}) constitutes a constraint that ensures the stationarity of sequence $\{x_{i\tau}| i\in \mathbb{Z}\}$.  The key of this constraint is $d_\tau \in (-2,0)$. This is because substituting $G_{\tau,i}$ in Eq.(\ref{eq:x_int}) yields  a first order  auto-regressive (AR(1)) process  with  process parameter $1+d_\tau$ driven by a stationary $f_{\tau,i}$, which is not necessarily white.  An AR(1) process is stationary when the magnitude of the process parameter is smaller than one (see e.g. \cite{SZ:99}), which is ensured by $d_\tau \in (-2,0)$.
Thus,  with $G_{\tau,i}$ in Eq. (\ref{eq:g_df}),  all sequences $\{x_{i\tau}| i\in \mathbb{Z}\}$ with all non-zero values of $\tau$ are  stationary. 

Substituting Eq.(\ref{eq:g_df}) into Eq.(\ref{eq:x_int}), yields,  after some manipulations, 
\begin{equation}
\begin{split}
       \sigma_{f_\tau}^2 = & ~~\sigma^2 \Big( 1-(1+d_\tau)^2 \Big ) \\
       =& -\sigma^2 d_\tau  \big ( 2+d_\tau\big )~~~\text{for}~\tau \in \mathbb{Z}_+ 
\end{split}
\label{eq:df-curve}
\end{equation}
where $\sigma_{f_\tau}^2$ is the variance of the fluctuating forcing $f_\tau$, again defined as a converging time average, like the variance  $\sigma^2$ in Eq.(\ref{eq:equilibrium_var}). 
Eq.(\ref{eq:df-curve}) results from the IFDR embodied in $G_{\tau,i}$. It states that the stronger the dissipation, the stronger is the fluctuation and with that  the larger the variance of $f_\tau$, $\sigma_{f_\tau}^2$.  When the strongest dissipation is reached for $d_\tau=-1$,  the fluctuating forcing $f_\tau$ is strongest,  characterized by the maximum variance  of $\sigma^2_{f_\tau}=\sigma^2$. 

We further postulate that as $\tau$ increases, there is a general tendency toward the strongest dissipation.  Once the strongest dissipation is reached for a certain value of $\tau$, say $\tau_o$, we have
\begin{equation}
\label{eq:df_tauo}
d_\tau=-1, ~ \sigma^2_{f_\tau}=\sigma^2_{f_{\tau_o}}, ~~~ \text{for all}~ \tau>\tau_o.
\end{equation} 
$G_{\tau,i}$ with $\tau>\tau_o$ equals then $c_\tau-x_i+f_{\tau,i}$.
For a sequence $\{f_{\tau,i\tau}|i\in\mathbb{Z}\}$ with $\tau>\tau_o$, $\gamma_{f_{\tau},k}$, the covariance between any two  members of this sequence, $f_{\tau,i\tau}$ and $f_{\tau,(i+k)\tau}$,  satisfies
\begin{equation}     
\gamma_{f_{\tau},k} = 
\begin{cases}
\sigma^2_{f_\tau}, & ~~~\textnormal{for}~  k=0, \\
 0, & ~~~\textnormal{for}~  k\neq 0 . 
\end{cases}
\label{eq:g_unified_f}
\end{equation} 
Again,  $\gamma_{f_{\tau},k}$ is  defined 
as a converging time average, similar to 
 $\gamma_\tau$  in Eq.(\ref{eq:acf}).  
According to Eq.(\ref{eq:g_unified_f}), the fluctuating forcing in $G_{\tau}$, $f_{\tau}$, behaves like a white noise for $\tau>\tau_o$.  We come back to the existence of $\tau_o$ in Section \ref{s10.7}.

Generally, the three quantities, $\sigma^2_{f_\tau}$, $\sigma^2$, and $d_\tau$, satisfy Eq.(\ref{eq:df-curve}) for  any stationary solution, whether it is generated by a forced-dissipative system in dynamical equilibrium or by a purely periodic and frictionless system.
However,  the tendency toward the strongest dissipation exists only for the former, not for the latter (s. also Fig.8 in \cite{jvonstorch:24}. 

A justification of the postulates requires long sequences of  $G_{\tau,i}$ for different values of $\tau$ and for all components of ${\bf x}$, and each of these sequences requires a record of the full solution of ${\bf x}$ of the considered system. This is out of reach for a high-dimensional dynamical system, such as a climate model,  not mentioning a many-particle system considered in statistical mechanics.  
However, the Lorenz 1963 model, which produces  equilibrium fluctuations as expected from a dynamical system in equilibrium with a reservoir,  can be used to validate our postulates. 

\begin{figure}[bt!]
\begin{center}
    \includegraphics[width=0.8\textwidth]{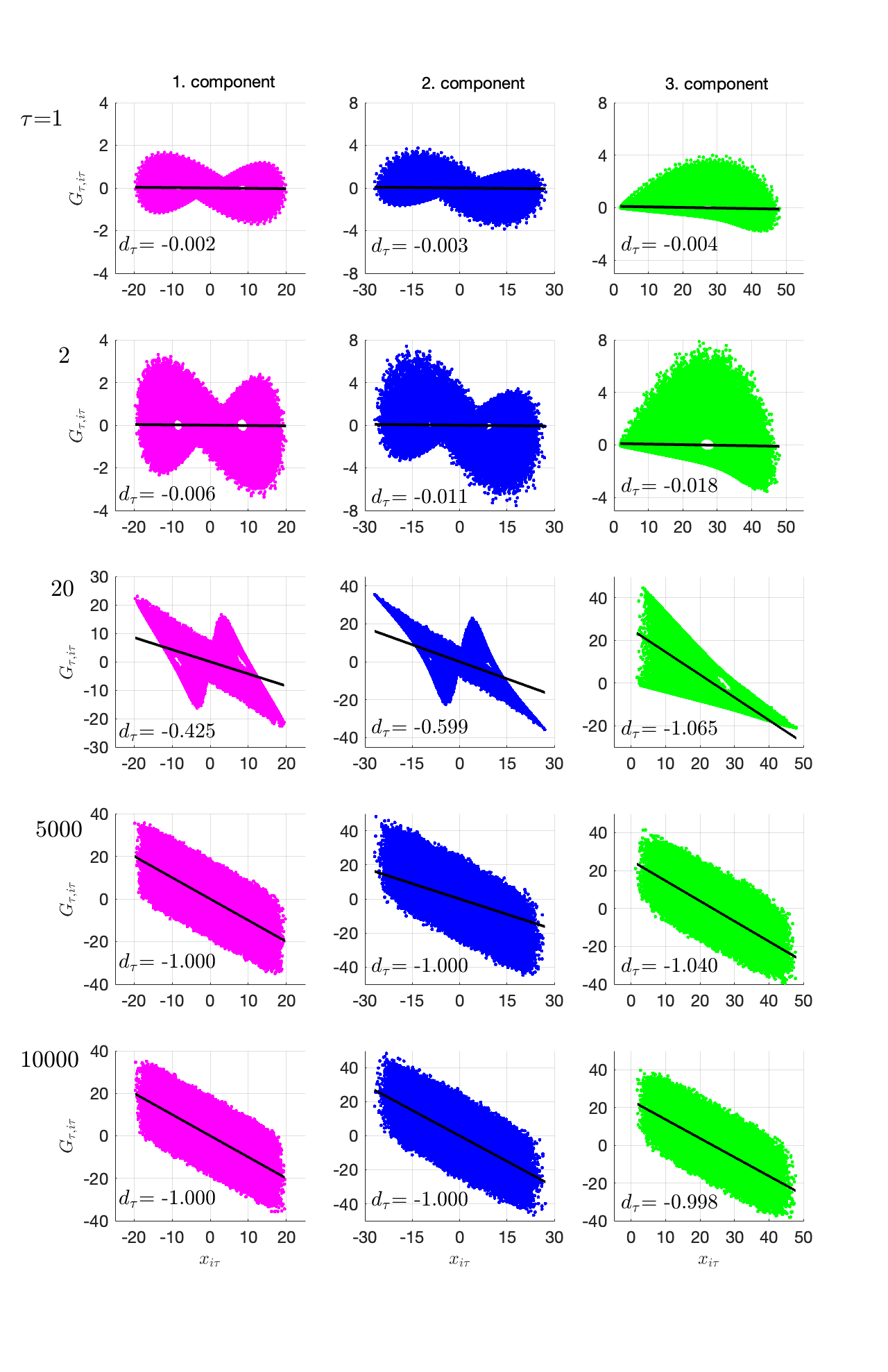} 
\caption{Scatter diagrams  of $G_{\tau,k}$  against $x_{k}$ (dots) with $k=i\tau$ and $i=1,2,\cdots, 10^5$, including the associated regression line (black) and the value of the regression slope $d_\tau$, for $\tau=$1, 20, 5000, 10000, and for the three Lorenz components (magenta, blue, green).  The slopes are  derived from $10^7$ pairs of $(x_k, G_{\tau,k})$ collected  along a stationary Lorenz solution. 
The Lorenz solution is obtained by integrating the Lorenz model from an equilibrated  state using a time step of 0.01.
}
\label{fig:g_x_regression}
\end{center}
\end{figure}

\begin{figure}[hbt!]
\begin{center}
    \includegraphics[width=1\textwidth]{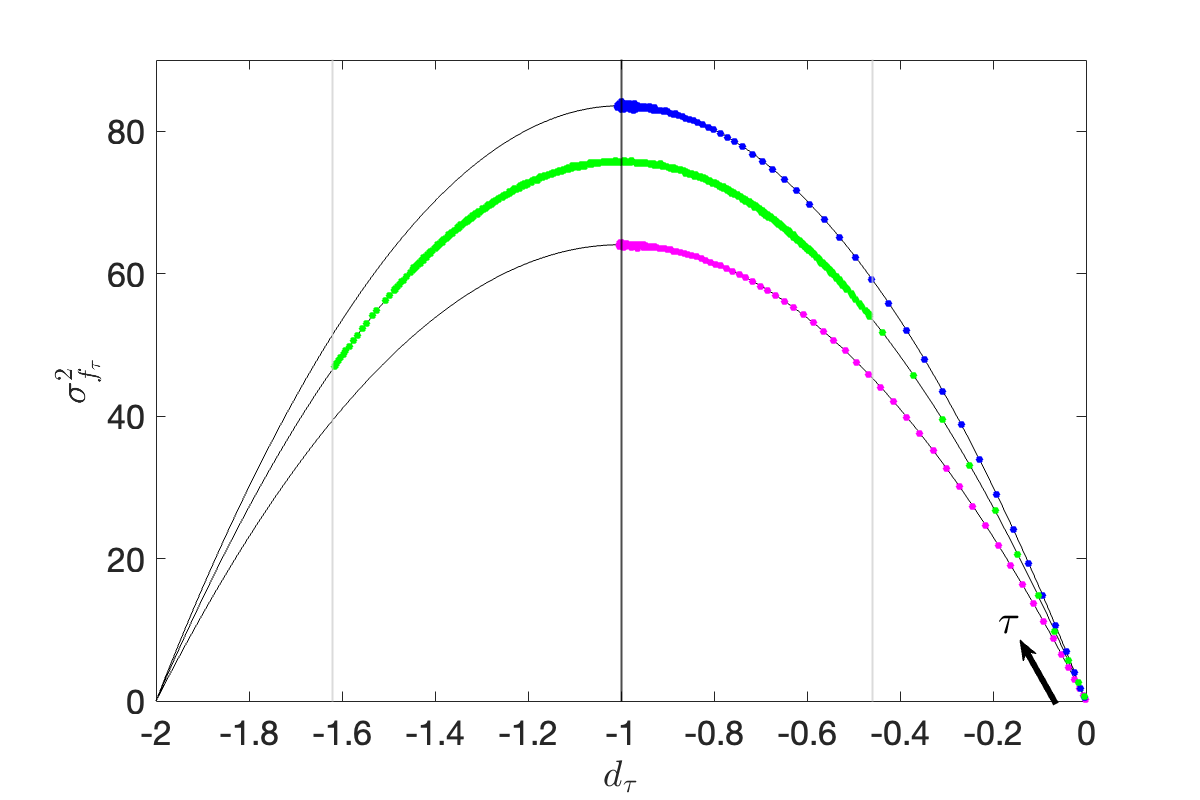} 
\caption{Relations between  $d_\tau$  and the variance of fluctuating component $f_{\tau,k}$, $\sigma^2_{f_\tau}$,  for $\tau=1,2,\cdots,1000$, and for the three Lorenz components (magenta, blue, green dots). Each point $(d_\tau,\sigma^2_{f_\tau})$  is obtained by regressing $G_{\tau,k}$ against $x_{k}$ using $(x_{k}, G_{\tau,k})$ with $k=i\tau$ and $i=1,2,\cdots, 10^6$ collected  along a stationary Lorenz solution, with $\sigma^2_{f_\tau}$ being the variance of the residual not represented by the regression. 
The black curves  are defined by Eq.(\ref{eq:df-curve}). The arrow indicates the direction in which the point $(d_\tau,\sigma^2_{f_\tau})$ moves with increasing $\tau$, starting from $\tau=1$. The left gray line marks $d_\tau=-1.62$, first reached by the green dot as $\tau$ increases; the right line, 
$d_\tau=-0.46$, shows its far most right position reached when further increasing $\tau$.  They show that the green dot approaches the point with the strongest dissipation at $d_\tau=-1$ in an oscillatory manner.
}
\label{fig:fd_curve}
\end{center}
\end{figure}

\begin{figure}[hbt!]
\begin{center}
    \includegraphics[width=1\textwidth]{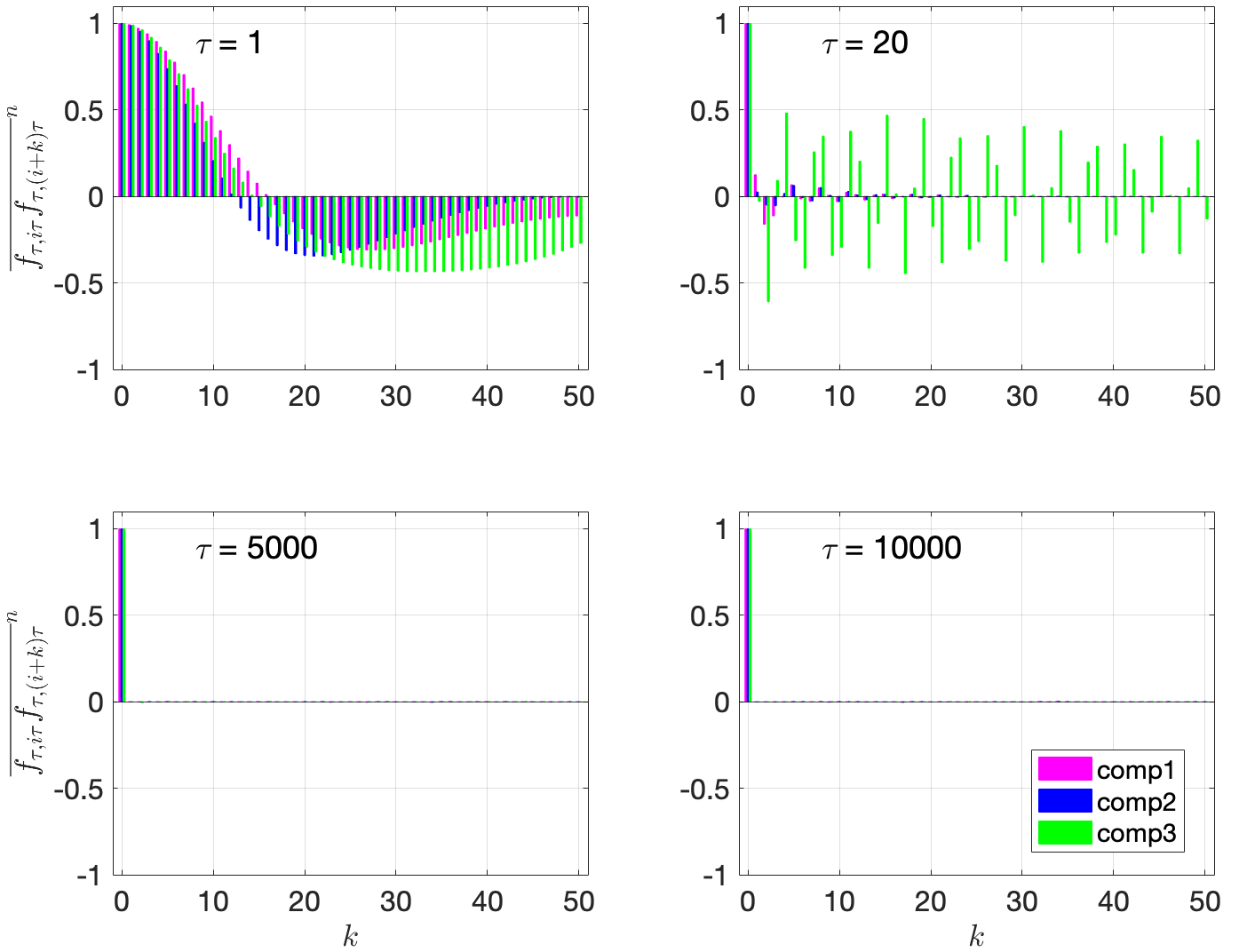} 
\caption{Auto-correlation functions of a series $\{f_{\tau,k}|k=i\tau, i\in\mathbb{Z}_+\}$,  $\overline{f_{\tau,i\tau}f_{\tau,(i+k)\tau}}^n$, as functions of  $k$, for $\tau$=1, 20, 5000, and 10000, derived by averaging over $n=10^6$ time points, for the three Lorenz components (magenta, blue and green). }
\label{fig:f_acf}
\end{center}
\end{figure}

Fig.\ref{fig:g_x_regression} shows for the Lorenz model that the regression slope $d_\tau$ is indeed negative and  in the interval $(-2, 0)$ for all considered values of $\tau$ and for all components $x$ of ${\bf x}$. For  sufficiently large values of $\tau$ for which $d_\tau$ is close or equal to -1 (third and fourth rows), the scatter approaches a parallelogram, indicating the variance of $f_\tau$, which is characterized by the width of the scatter around the regression line, becomes independent of $x$, consistent with Eq.(\ref{eq:df_tauo}). Given $d_{\tau_o}=-1$, $\sigma^2_{f_{\tau_o}}$ equals, according to Eq.(\ref{eq:df-curve}), the variance of $x$, $\sigma^2$. 

Fig.\ref{fig:fd_curve}  shows that  $d_\tau$ and the variance of $f_\tau$, $\sigma_{f_\tau}^2$, are indeed related to each other according to Eq.(\ref{eq:df-curve}).
More importantly, Fig.\ref{fig:fd_curve} also illustrates the general tendency toward the strongest dissipation and maximum variance of $f_\tau$, marked by the center point at  $(d_\tau,\sigma_{f_\tau}^2)=(-1,\sigma^2_{f_{\tau_o}})=(-1,\sigma^2)$,  followed by the halt at this point as $\tau$ further increases. 
For the first and the second Lorenz component (magenta and blue dots), this tendency is monotonous. It starts from the point $(d_\tau,\sigma^2_{f_\tau}) \simeq (0,0)$ at the far right for  $\tau=1$, moves  toward the  center where $d_\tau=-1$ as $\tau$ increases, and stays  at the center as $\tau$ further increases.  For the third component (green dots), this tendency is not monotonous.  It  also starts  from  $(d_\tau,\sigma^2_{f_\tau}) \simeq (0,0)$ for  $\tau=1$. However, as $\tau$ increases,  $(d_\tau,\sigma^2_{f_\tau})$ swings  around the center,  thereby gradually approaches the center, and eventually reaches and stays at the center for sufficiently large $\tau$ (more visible from Fig.5 in \cite{jvonstorch:24}). Thus, in all three cases, when $d_\tau$ converges to -1 with increasing value of $\tau$, $\sigma^2_{f}$  converges to its maximum $\sigma^2_{f_{\tau_o}}=\sigma^2$. 

Fig.\ref{fig:f_acf} shows that for sufficiently large  $\tau$, for which $d_\tau$ is about -1,  the fluctuating forcing $f_{\tau,i}$ has an auto-correlation function visually identical to that given in Eq.(\ref{eq:g_unified_f}).  Thus, the Lorenz solutions confirm our postulates about the properties of $G_\tau$  (for more details see \cite{jvonstorch:24}).

With the IFDR formulated in Eq.(\ref{eq:g_df}), we state that only systems governed by those $F({\bf x})$, whose integrals take the form of IFDR given in Eq.(\ref{eq:g_df}), are able to  attain equilibrium fluctuations in $x$, with  their characteristics described by Section \ref{s10.2}. This peculiar constraint on $F$, which  is most likely attainable for  a dissipative system subjected to a constant and reservoir-like external forcing, is upgraded as one of the two principles governing fluctuations in dynamical equilibrium  \citep{vonstorch2026}. 

\section{Dissipation in $G_\tau$ versus dissipation in $F$}\label{s10.7}

This section shows that the dissipation of $x$ inherent in its  integral forcing $G_\tau$---also referred to as the \IND{total dissipation} of $x$---differs from that encoded in the differential forcing $F({\bf x})$ of $x$. For simplicity, we assume that the latter is linear and given by $-\alpha x$ with some positive constant $\alpha$, although the conclusions below do not depend on this specific form. The remainder of $F({\bf x})$---denoted as the interaction term $H({\bf x})\equiv F({\bf x})+\alpha x$---represents the interaction of $x$ with the other components $x'$ of ${\bf x}$ (and the influence of the constant external forcing).  Eq.(\ref{eq:x}) reduces then to
\begin{equation}\label{eq:x_H}
    \frac{\mathrm{d}x}{\mathrm{d}t}=-\alpha x +H({\bf x}(t)).
\end{equation}
When time is discretized using increment $\delta t$,  integrating Eq.(\ref{eq:x_H}) yields
\begin{equation}\label{eq:x_GH_disc} 
x_{i+\tau}  = (1- \alpha ~\delta t)^{\tau}x_i +J_{\tau,i},  ~~~\tau\in\mathbb{Z}_+ 
\end{equation}
where $J_{\tau,i}$ is defined  as
\begin{equation}\label{eq:J}
    J_{\tau,i} = \sum_{j=1}^{\tau} (1-\alpha ~\delta t)^{\tau-j}~ H_{i+j-1} ~\delta t, ~~~\tau\in\mathbb{Z}_+, 
\end{equation}
where $H_i=H({\bf x}_i)$.
The two subscripts of $J_{\tau,i}$ have the same meaning as those of $G_{\tau,i}$.  
 
After some manipulations,   we obtain from Eq.(\ref{eq:x_GH_disc})   the  total dissipation of $x$ within $\tau$ time steps
\begin{equation}\label{eq:d_fH}
   d_\tau  = (1-\alpha ~\delta t)^\tau  + \gamma_{\tau}^{xJ} / \sigma^2  -1 , 
\end{equation}
where $\gamma_\tau^{xJ}$ is the cross-covariance  between $x_i$ and $J_{\tau,i}$ and given by
\begin{equation}\label{eq:gamma_xJ}
\gamma_{\tau}^{xJ}  = \lim_{n\rightarrow \infty} \frac{1}{n}\sum_{i=1}^n  \big(x_i-\overline{x}^n \big)J_{\tau,i}.
\end{equation}

Thus, both the dissipation term $-\alpha x$ and  an integral of the interactive term $H$ contribute to  $d_\tau$.  However, they do so in fundamentally different ways.  The  contribution $(1-\alpha ~\delta t)^\tau$ from $-\alpha x$ is evident and effective {\em without any integration}. The contribution $\gamma^{xJ}_\tau/\sigma^2-1$ from an integral of $H$,  on the other hand,  arises from the correlation between $x_i$ and $H_{i+j-1}$ (with $j=1,\cdots,\tau$), or via the dependence of $H$ on components $x'\neq x$,  from the  correlation between $x_i$ and $x'_{i+j-1}$ (with $j=1,\cdots,\tau$). The latter weakens with increasing $j$, due to the dissipation terms in the differential forcings of  $x'$ and $x$.  
This loss of correlation,  which is related to the dissipation of $x'$ and hence ``not visible'' in $F$ of $x$,  becomes only apparent {\em after integrating the full system forward in time}. 

Because the interaction between $x$ and $x'\neq x$ requires time, the dissipative term in the differential  forcing of $x'$ cannot be recognized by $x$  over sufficiently short integration times, making $H$ act as a driver. This driving can persist with increasing integration time. However, the effect from $H$, no matter whether it drives or damps,  eventually vanishes due to the dissipation terms in equations of all components of ${\bf x}$ that interact with $x$. As the effect of these dissipation terms unfold,  the total dissipation  reaches its strongest strength characterized by $d_\tau=-1$.   

Below, we illustrate this by considering  the contributions to $\rho_\tau=1+d_\tau$ from $-\alpha x$ and $\mathcal{H}$  derived from   the Lorenz 1963 model. Each of the three Lorenz equations contains a linear dissipation  with  $\alpha=$ 10, 1, and 8/3, respectively. When denoting the three components by $u$, $v$ and $w$, the interactive term $H({\bf x})$ equals $10v$, $u(28-w)$, and $uv$ for the three components, respectively. 

\begin{figure}[hbt!]
\begin{center}
    \includegraphics[width=0.95\textwidth]{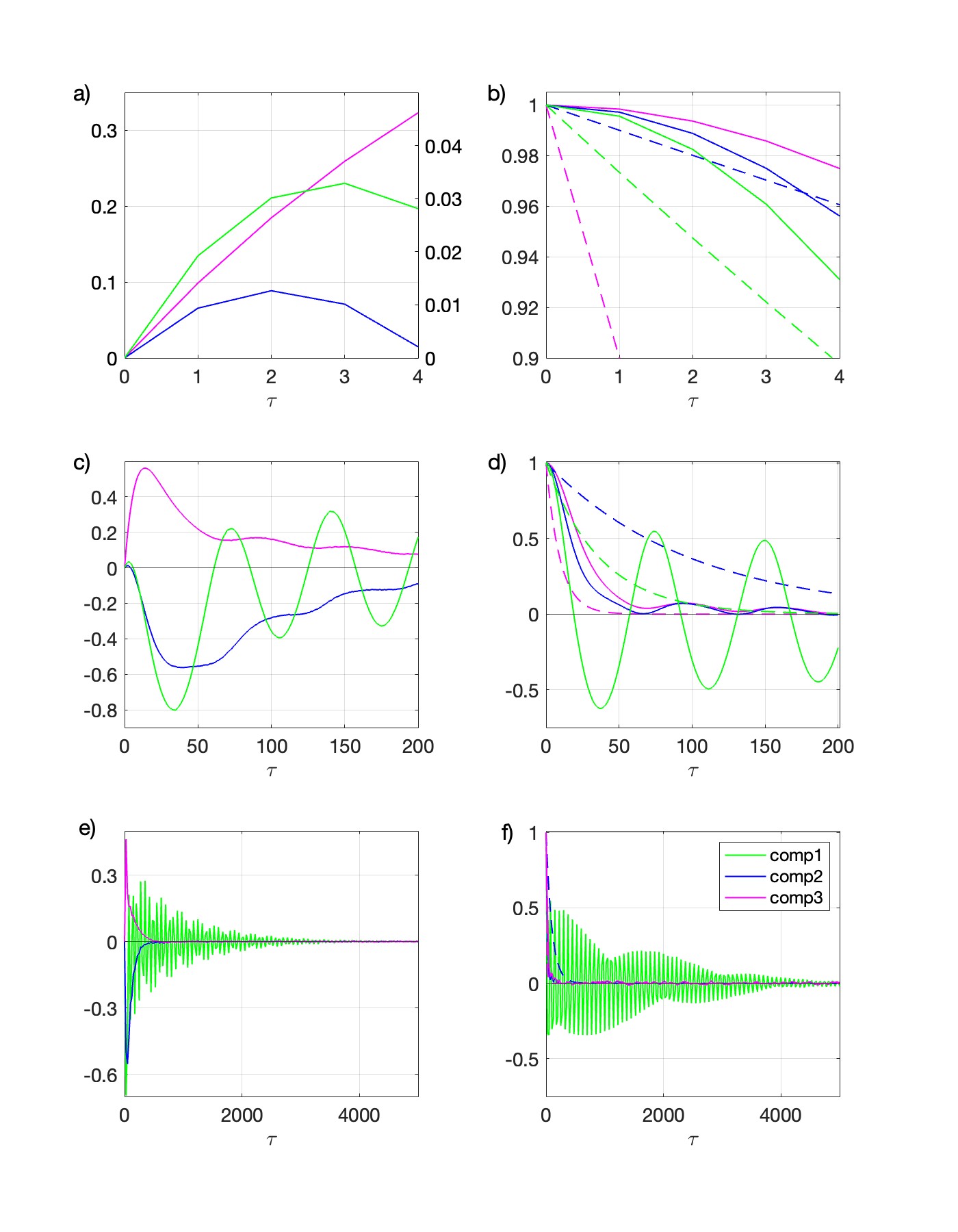} 
\caption{Normalized covariance function $\gamma^J_\tau/\sigma^2$ in the left panel, and  auto-correlation function $\rho_\tau$ (solid lines) and the instantaneous damping $(1-\alpha\delta t)^\tau$ (dashed lines)  in the right panel. The top, middle and bottom rows shown these functions for $\tau=0,1,3,4$,   $\tau=0,1,\cdots, 200$, and $\tau=1,21,41,\cdots, 3981$, respectively.   The colors (magenta, blue and green) indicate the  three Lorenz components.    The correlations are derived from stationary Lorenz solutions of length $n=10^6$. In a), the left y-axis is for $\gamma^J_\tau/\sigma^2$ derived from the first, and the right y-axis for $\gamma^J_\tau/\sigma^2$  derived from the second and the third  Lorenz component.  }
\label{fig:cvar}
\end{center}
\end{figure}

For small values of $\tau$, $\gamma^{xJ}_\tau/\sigma^2$ is positive (Fig.\ref{fig:cvar}a).  According to Eq.(\ref{eq:d_fH}), a positive $\gamma^{xJ}_\tau/\sigma^2$ produces, in case $\alpha=0$, a positive $d_\tau$. Such a $\gamma^{xJ}_\tau/\sigma^2$ acts   as a driver. The driving effect counteracts  the exponential decay  caused by $-\alpha x$ (Fig.\ref{fig:cvar}b), making the decay of  $\rho_\tau$ (solid lines) slower than that of $(1-\alpha ~\delta t)^\tau$ (dashed lines).    

To explore this driving effect  for even smaller integration time, we consider $d_1$, the total dissipation over one time step of length $\delta t$, and infer from that  the influence of  $\gamma^{xJ}_1/\sigma^2$  on $d_1$. Fig.~\ref{fig:d1}a) shows  that the driving effect of $\gamma^{xJ}_1/\sigma^2$ makes $|d_1|$ smaller than  $\alpha \delta t$, the effect of the dissipative term in $F$  within one time step. Moreover, as $\delta t$ decreases,  $d_1$ decreases proportionally to $-\delta t^2$ (solid lines), in contrast to the proportionality of $-\delta t$ associated with the dissipative term in $F$ (dashed lines). Thus, in the limit $\delta t \rightarrow 0$, while the rate of dissipative term in $F$  equals $-\alpha \neq 0$,   the rate of total dissipation associated with $G_1$ vanishes:
\begin{equation}\label{eq:limit_d1}
    \lim_{\delta t\rightarrow 0} \frac{d_1}{\delta t} =0,
\end{equation}
leaving $F({\bf x})$ effectively non-dissipative. 

\begin{figure}[hbt!]
\begin{center}
    \includegraphics[width=1\textwidth]{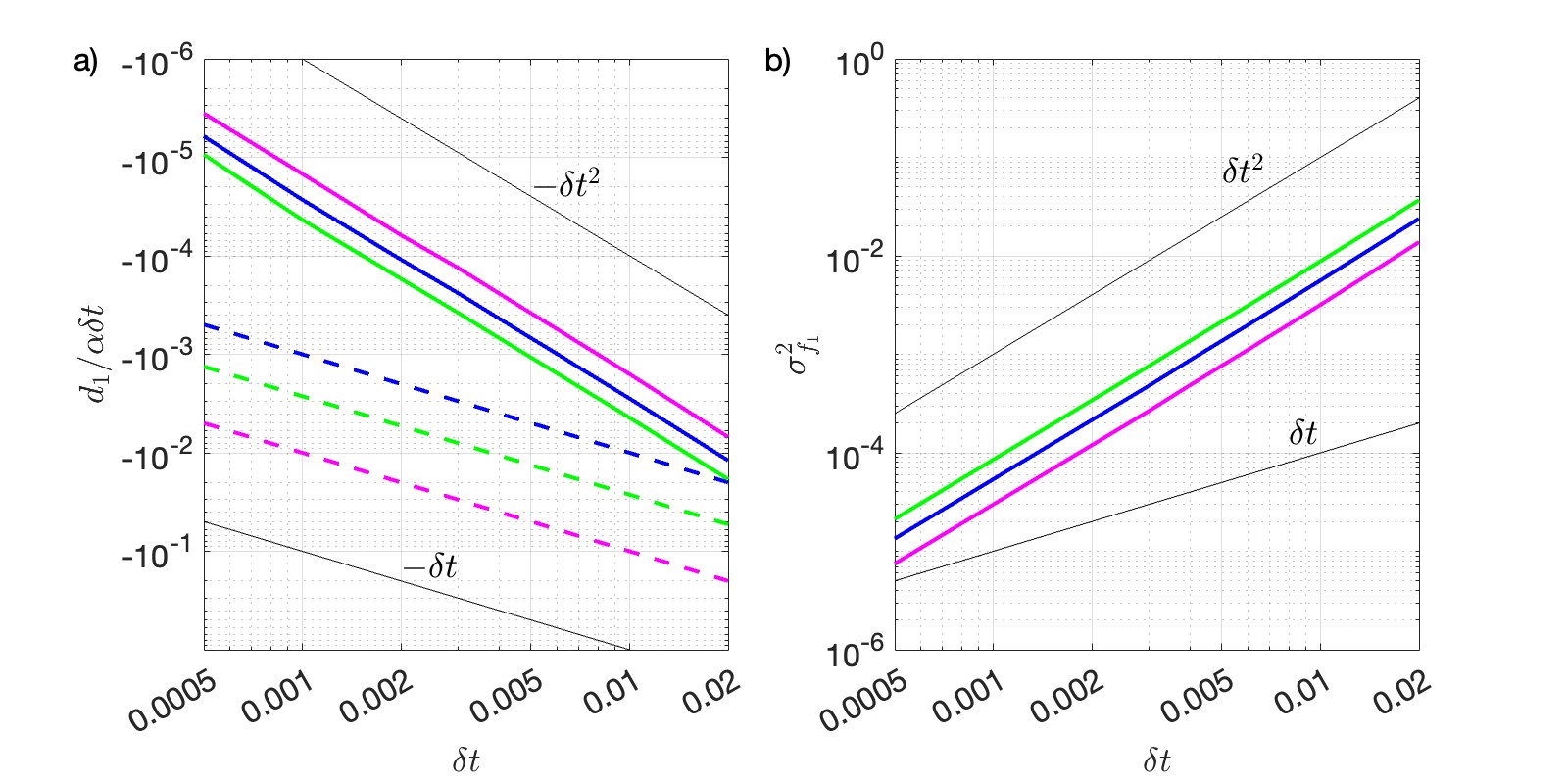} 
\caption{a) Total dissipation within one time step $d_1$ (solid lines),   the instantaneous damping within one time step $\alpha \delta t$ (dashed lines), and b) the variance of the corresponding fluctuating forcing $f_1$, $\sigma^2_{f_1}$, all as functions of the length of time step $\delta t$ for the three Lorenz components (magenta, blue, and green).  The two black lines  are, respectively, proportional to $-\delta t$ and $-\delta t^2$ in a), and to $\delta t$ and $\delta t^2$ in b).}
\label{fig:d1}
\end{center}
\end{figure}

For moderate values of $\tau$ (Fig.\ref{fig:cvar}c), $\gamma^{xJ}_\tau/\sigma^2$  is positive for the first Lorenz component (magenta), negative for the second (blue), and oscillatory for the third (green).   The driving effect of $H$ is evident for the first component, but also visible for the third component, for which $\gamma^{xJ}_\tau/\sigma^2$ is larger at $\tau=140$ relative to that at $\tau=73$.   
Due to the contribution $|\gamma^{xJ}_\tau/\sigma^2|$,  the decay of $\rho_\tau$ is  noticeably distinct from that predicted by $-\alpha x$ (solid versus dashed lines in Fig.\ref{fig:cvar}d).

For large values of $\tau$ (Fig.~\ref{fig:cvar}e,f), influenced not only by the dissipative term in $F$ of $x$ but also that of $x'$,   both $(1-\alpha \delta t)^\tau$ and $|\gamma^{xJ}_\tau/\sigma^2|$   decrease with increasing $\tau$. As  the strongest dissipation emerges,   $\rho_\tau$ decays to zero.  

Mathematically, the exponential function $(1-\alpha \delta t)^\tau$ indicates that the dissipative term in $F$ vanishes only in the limit $\tau \rightarrow \infty$. Physically, however, it is reasonable to treat the vanishing as effectively completed at some finite $\tau$. This is because, when compared to the infinite timescale over which a dissipative system is sustained by a reservoir---where ``infinite'' reflects the reservoir’s unlimited capacity---any real dissipative process should be considered as operating over a finite timescale. In the context of Eq. (\ref{eq:gamma0}), this implies that the summation’s lower and upper bound correspond to a ``greater infinity'' than any $\tau$ achievable by a non-zero $\rho_\tau$.
By the same argument, the timescale associated with any physical process included in  differential forcings of $x'\neq x$ should likewise be considered as finite.

In practice, estimating correlations at large time lags is difficult because a finite-length solution provides only a limited number of data points. This limitation can be mitigated by using an ensemble of solutions. Such an ensemble naturally arises from the fact that a system in dynamical equilibrium with a reservoir admits many macroscopically equivalent realizations---each characterized by the same macroscopic quantities. 
Here, the ensemble is only used  for generating more data points, regardless of its associated probability distribution function. When the number of data points is increased by  using an ensemble, correlations at sufficiently large time lags become visually indistinguishable from zero \citep{JvS:22}. 

That all processes in a physical system operate on finite timescales---when compared with  the infinite time horizon  of the external forcing---implies the existence of a finite threshold  $\tau_o$ such that for $\tau>\tau_o$,  $d_\tau=-1$ and $\rho_\tau=0$. For such large values of $\tau$,  $G_{\tau,i}$ reduces to $c_\tau-x_i+f_{\tau,i}$ and Eq.(\ref{eq:x_int}) to 
\begin{equation}
    x_{i+\tau}=c_\tau+f_{\tau,i} ~~~\text{for}~~ \tau>\tau_o.
\end{equation}
The solution  $\tau>\tau_o$ time steps later is determined by the fluctuating forcing $f_{\tau,i}$ in $G_{\tau,i}$, independent of   the initial state $x_i$. Such a solution is random and irreversible. 

Before proceeding further, we consider the special case when $\gamma^{xJ}_\tau$ vanishes.  This happens when the interactive term $H$ of $x$ is a function of components $x'\neq x$, whose values, $H_{i+j-1}$ with $j=1,\cdots,\tau$,  are however independent of  $x_i$. The stochastic treatment of Brownian motion adopts this assumption: In the velocity equation of a Brownian particle,   the particle’s velocity at a time   is considered  independent of  variables of impacting molecules at that and after that time. Under this assumption,  $H$ can be represented by a stochastic forcing, leading to $\gamma^{xJ}_\tau=0$ for all $\tau\geq 1$, also in the limit $\delta t \rightarrow 0$. The problem associated with  this special case  will be discussed  in the next section.

Generally, physical systems of our interest---such as climate models---do not satisfy the condition $\gamma^{xJ}_\tau=0$ for all $\tau\geq 1$.
 For a solution of a variable $x$ simulated by such a model, the contribution of $\gamma^{xJ}_\tau$ to the variable’s autocorrelation function $\rho_\tau$  is typically non-negligible. Consequently, the decay of $\rho_\tau$   cannot be attributed solely to the explicit damping term (e.g., $-\alpha x$) in the governing equation of that variable.

\section{The limit {$\delta t \rightarrow 0$}}\label{s10.8}

In the spirit of IFDR, the decrease of $d_1$ with decreasing $\delta t$ must be accompanied by a similar decrease of $\sigma^2_{f_\tau}$. Fig.\ref{fig:d1}b shows that this is indeed the case for the Lorenz model.   As $\delta t$ decreases, the magnitudes of both $d_1$ and $\sigma^2_{f_1}$ decrease in tandem, each scaled with $\delta t^2$. 
Thus, not only the IFDR, but also the rate of change of IFDR,  vanishes in the limit $\delta t \rightarrow 0$. 
Before discussing the implication for equilibrium fluctuations   in the next section, we point out below that a $\delta t^2$-scaling results from the dependence of  the interaction term $H$ of $x$ with  $x$ itself. 

We consider the Brownian motion, with $x$ representing the  velocity of a Brownian particle. If this velocity is assumed independent of the impinging water molecules,  $H({\bf x})$ would be independent of $x$. It would be perfectly meaningful to describe $H({\bf x})$ as a stochastic  forcing  $\zeta$. Formally, an integral of $\zeta$ over $\delta t$ is described by  a Wiener process $dW$. It is known from mathematical consideration that the variance of the Wiener process $dW$  scales with $\delta t$. Thus, it is the independence of $H$ with $x$, that  makes the scaling of an integral of $H$ over $\delta t$ to deviate from $\delta t^2$.

However, the assumption that $H({\bf x})$ is independent of $x$, which underpins a stochastic description, is inconsistent with our understanding of the dynamics governing Brownian velocity. For a sufficiently short time step $\delta t$---during which the Brownian particle collides with only a single water molecule, a condition that is always attainable since time is continuous---momentum is exchanged between the particle and the molecule. This exchange renders the particle’s velocity dependent on that of the molecule, resulting in a $\delta t^2$-scaling and a nonzero $\gamma^{x\mathcal{H}}_\tau$. We therefore conclude that, in the limit $\delta t \rightarrow 0$, the interaction term $H({\bf x})$ cannot be independent of $x$.

\section{Complementary yet distinct spheres of $F$ and  $G_\tau$}\label{s10.9}

In the limit $\delta t \rightarrow 0$, the rate of change of IFDR,  as characterized by $d_1/\delta t$ and $\sigma^2_{f_1}/\delta t$, disappears  faster than the convergence  of the terms in the differential equations, making the latter as the only governing principle. The effect of $G_\tau$ is hence limited.  This  reminds us of the other limitation---the one associated with $F$  discussed in Section \ref{s10.4}. 

The respective limitations of $G_\tau$ and $F $ highlight that equilibrium fluctuations must be understood, in their entirety, as being governed by two complementary principles that cannot be reduced to one another: the differential equations defined by $F$ and the IFDR embodied in $G_\tau$   \citep{vonstorch2026}. Here, “in their entirety” refers to all aspects of equilibrium fluctuations, encompassing both microscopic and macroscopic levels. The microscopic description is provided by individual solutions of the dynamical equations, whereas the macroscopic description is given by macroscopic quantities defined as time averages that converge in the long-time limit.

These two principles are complementary in the sense that one is required to determine the system’s solutions, while the other is necessary to determine its macroscopic properties. At the same time, they are irreducible to one another: the IFDR encoded in $G_\tau$  does not exist as a time derivative and therefore cannot be incorporated into the differential forcing $F$.
The additional information contained in the IFDR---absent from the differential equations---arises from cross-component interactions and manifests itself as a dissipation whose full effect only unfolds when the system is integrated forward in time. This integrated effect is culminated in the quantity $\Gamma^x(0)$, which cannot be inferred from $F$, not even from the spectral properties of $F$.

The IFDR enforces  a set of constraints, each defined for an integration time over $\tau$  steps---with $\tau\in\mathbb{Z}_+$---to ensure the  stationarity of a solution sampled at every $\tau$ steps.  The  agglomerate of these constraints is necessary for the stationarity of the whole solution.  The latter is the guarantor for time averages to converge with increasing averaging time.  As integration time goes to zero, nothing is integrated. The IFDR, which describes two effects of  integrated $F$, disappears, and it disappears at the rate of $\delta t^2$.

The IFDR embodied in $G_\tau$ of $x$ should be interpreted as the cause of all \IND{macroscopic quantities} related to $x$, since such a macroscopic quantity is a function of the solution of $x$ and since the solution of $x$ at a future time  is (given an initial state) determined and only determined by its respective integral forcing $G_\tau$. That all macroscopic quantities  result from  the IFDR embodied in $G_\tau$ is further substantiated below by  expressing them as functions of $d_\tau$ and $\sigma^2_{f_\tau}$, the two quantities  describing the strengths of dissipation and fluctuations in $G_\tau$. For the total variance  of $x$, $\sigma^2$, we have from Eq.(\ref{eq:df_tauo}) and Eq.(\ref{eq:df-curve}) 
\begin{equation}
    \sigma^2=\sigma^2_{f_{\tau_o}}.
\end{equation}
For the  auto-correlation $\rho_\tau$, its relation to $d_\tau$ is stated in Eq.(\ref{eq:d_rho}).  For  the auto-covariance $\gamma_\tau=\rho_\tau \sigma^2$, its relation to   $d_\tau$ and $\sigma^2_{f_\tau}$  follows from Eq.(\ref{eq:df-curve}) and reads
\begin{equation} \label{eq:gamma}
    \gamma_\tau = \frac{1+d_\tau}{1-(1+d_\tau)^2} ~ \sigma^2_{f_\tau} 
    = - \frac{\sigma^2_{f_\tau}}{d_\tau} ~\Bigg ( \frac{1}{\frac{1}{1+d_\tau}+1} \Bigg) ~~~~\text{for } \tau\in\mathbb{Z}_+.
\end{equation}
To derive the lag-zero auto-covariance $\gamma_0$, we first set $\tau=1$ in Eq.(\ref{eq:df-curve})  and derive the ratio $\frac{\sigma^2_{f_0}}{-d_0}$ as $\frac{\sigma^2_{f_1}}{-d_1}$ in the limit $\delta t \rightarrow 0$:
\begin{equation}
    \frac{\sigma^2_{f_0}}{-d_0} \equiv \lim_{\delta t \rightarrow 0}
\frac{\sigma^2_{f_1}}{-d_1}=\lim_{\delta t \rightarrow 0} \sigma^2 \big( 2+d_1\big)=2\sigma^2=2\sigma^2_{f_{\tau_o}},
\end{equation}
where $\lim_{\delta t \rightarrow 0}d_1=0$ and the result $\sigma_{f_o}^2=\sigma^2$ are used.
Defining then $\gamma_0$ as $\gamma_1$ given in Eq.(\ref{eq:gamma}) in the limit $\delta t\rightarrow 0$  yields 
\begin{equation}
    \gamma_0\equiv\lim_{\delta t\rightarrow 0} -\frac{\sigma^2_{f_1}}{d_1} ~\Bigg ( \frac{1}{\frac{1}{1+d_1}+1} \Bigg) 
    =\sigma^2_{f_{\tau_o}}.
    \label{eq:gamma_null}
\end{equation}
Using Eq.(\ref{eq:gamma}) and Eq.(\ref{eq:gamma_null}), and the equality $\rho_\tau=\rho_{-\tau}$, \IND{$\Gamma^x(0)$} associated with a continuous solution of $x$ is given by
\begin{equation}
\Gamma^x(0)=\lim_{\delta t \rightarrow 0}\sum_{\tau=1}^{\infty} - \frac{\sigma^2_{f_\tau}}{d_\tau} ~\Bigg ( \frac{1}{\frac{1}{1+d_\tau}+1} \Bigg) +\sigma^2_{f_{\tau_o}}.
\end{equation}

In the above consideration, $\gamma_0$  is intentionally first derived for the discrete case, then extended to the continuous case via the limit $\delta t\rightarrow 0$. This procedure is necessary since the time rate of changes associated with $d_1$ and $f_1$, $d_1/\delta t$ and $\sigma^2_{f_1}/\delta t$, vanish in the limit $\delta t \rightarrow 0$, and  is hence not representable in any differential equations,  highlighting that the principle for macroscopic quantities cannot be derived directly from the microscopic differential equations.


\section{Concluding remarks}\label{s10.10}

Connecting statistical mechanics with climate science and leveraging climate research methodologies to investigate fluctuating phenomena offer promising avenues for new insight. Drawing inspiration from the Langevin equation, Hasselmann explained variations in slow climate variables as the integrated effect of fast processes---a mechanism fundamentally different from the prevailing theories of the 1970s.  
Meanwhile, the numerical simulation of the climate system has become a cornerstone of climate science, providing extensive evidence that, in equilibrium (realized through control simulations), the variances of climate variables are well-defined. More importantly, direct numerical simulation offers a concrete pathway to link the well-established fluctuation–dissipation theorem to the underlying microscopic equations. This connection, however, has so far not been sufficiently explored, particularly in a formulation that avoids the use of any specific approximations. Exploring this link leads to the identification of the integral fluctuation dissipation relation (IFDR). 

The IFDR represents, in addition to the differential equations as one governing principle,  the other principle governing macroscopic properties of a particular class of forced-dissipative systems. Such a system possesses several distinctive features: it is described by a multi-dimensional state vector ${\bf x}$; its evolution  is governed by a set of differential equations in time, each defined by a differential forcing $F({\bf x})$ that includes some kind of damping; it is in a dynamical equilibrium with a constant external  forcing---such as that supplied by a reservoir. A defining signature of a dynamical equilibrium is the finite but nonzero spectral density of ${\bf x}$ at zero frequency.  The two principles are  irreducible to each other but interlinked in the sense  that   the IFDR resides in integrals of differential forcing $F({\bf x})$. 

The IFDR is a necessary condition for the existence of a dynamical equilibrium, analogous to the requirement that $F({\bf x})$ be the zero function in the case of a steady-state equilibrium. However, unlike the steady-state condition---which is a single constraint---the fluctuation-dissipation relation encompasses a family of conditions, each associated with a nonzero integration time.  In this role, the IFDR defines all macroscopic quantities characteristic for equilibrium fluctuations.

The IFDR  is only effective when the considered system is integrated forward in time. The need for forward integration is a consequence of the fact  that the total dissipation of a component $x$ of ${\bf x}$ arises not only from  the  dissipation included in the differential forcing $F$ of $x$ but also from that included in $F'$ of $x'\neq x$, which interacts with  $x$.  The latter only unfolds with  increasing integration time so that  the strongest dissipation of  $x$ is only reached when the dissipation in all components of ${\bf x}$, that influence $x$, has been unfolded after a long integration time. 

The need for two complementary yet distinct principles is a manifestation of the fact that equilibrium fluctuations  do not result from a simultaneous balance between generation and dissipation.  Over an infinitesimal $\delta t$, the interaction of $x$ with $x'\neq x$ acts as 
 a driver that dominates the dissipation of $x$,   making the  differential forcing $F$ of $x$ effectively non-dissipative,  thereby sustaining a continuous generation of fluctuations in $x$. This continuous generation is counteracted by the total dissipation of $x$, which reaches its strongest strength  when the system is integrated over a sufficiently long time period. The necessity of forward integration for  dissipation  to unfold makes the solution inherently unidirectional and irreversible.

The identification of IFDR offers a new perspective on the long-standing problem of the arrow of time. Existing approaches, most prominently embodied in fluctuation theorems \citep{evans2016}, have focused on characterizing irreversibility, while largely remaining agnostic about the origin of randomness. Since the emergence of randomness already introduces an intrinsic time asymmetry, a formulation that treats randomness as given may overlook an essential aspect of irreversibility.  The IFDR provides a natural framework in which irreversibility and randomness are treated on the same footing. From this viewpoint, reconsidering the arrow of time in light of IFDR appears both timely and potentially fruitful.

\noindent {\em Acknowledgment}:
I thank Klaus Hasselmann for his inspiration and constant encouragement. My thanks also go to Andreas Hense for carefully reading an early version of this essay and providing thoughtful comments. I am grateful for the opportunity to work at the Max Planck Institute for Meteorology, an institution that values and nurtures innovation and discovery.


\let\cleardoublepage\clearpage


\bibliographystyle{spbasic}
\bibliography{references}

\end{document}